\documentclass[12pt]{article}
\usepackage{amsmath,amsthm,amssymb,cite,epic,eepic,graphicx,lscape,subfigure,color}
\usepackage[width=142mm]{caption}
\begin{document}

\newcommand{\nc}{\newcommand}

\nc{\bea}{\begin{eqnarray}}
\nc{\eea}{\end{eqnarray}}

\nc{\nn}{\nonumber} 

\nc{\dxo}[1]{\frac{d{#1}}{dx_1}}
\nc{\dxt}[1]{\frac{d{#1}}{dx_2}}
\nc{\dy}[1]{\frac{d{#1}}{dy}}

\nc{\ddx}[1]{\frac{d^2{#1}}{dx^2}}
\nc{\ddy}[1]{\frac{d^2{#1}}{dy^2}}

\nc{\ddxy}[1]{\frac{d^2{#1}}{dxdy}}
\nc{\ddyx}[1]{\frac{d^2{#1}}{dydx}}

\nc{\re}[1]{\Re\mbox{e}(#1)}
\nc{\ddit}[1]{\frac{d^2{#1}}{dx_1dx_2}}
\nc{\ddto}[1]{\frac{d^2{#1}}{dx_2dx_1}}
\nc{\ddoo}[1]{\frac{d^2{#1}}{dx_1^2}}
\nc{\ddtt}[1]{\frac{d^2{#1}}{dx_2^2}}
\nc{\ddyy}[1]{\frac{d^2{#1}}{dx_2^2}}
\nc{\xx}{(x_1,x_2)}
\nc{\bxi}{{\bf \xi}}
\nc{\bxo}{{\bf \xi_1}}
\nc{\bxt}{{\bf \xi_2}}

\newtheorem{theorem}{Theorem}[section]
\newtheorem{lemma}[theorem]{Lemma}
\newtheorem{corollary}[theorem]{Corollary}
\newtheorem{definition}[theorem]{Definition}
\newtheorem{remark}[theorem]{Remark}
\newtheorem{remarks}[theorem]{Remarks}
\newtheorem{example}[theorem]{Example}
\nc{\bex}{\begin{example}}
\nc{\eex}{\end{example}}
%
%
\nc{\ga}{\alpha}
\nc{\gb}{\beta}
\nc{\gd}{\delta}
\nc{\gep}{\varepsilon}
\nc{\gz}{\zeta}
\nc{\gt}{\theta}
\nc{\gk}{\kappa}
\nc{\gl}{\lambda}
\nc{\gp}{\phi}
\nc{\gs}{\sigma}
\nc{\go}{\omega}
\nc{\gn}{\nu}
\nc{\gr}{\rho}


\nc{\cF}{\mathcal{F}}
\nc{\cP}{\mathcal{P}}
\nc{\cS}{\mathcal{S}}
\nc{\cN}{\mathcal{N}}
\nc{\cD}{\mathcal{D}}
\nc{\cH}{\mathcal{H}}
\nc{\cO}{\mathcal{O}}
\nc{\cT}{\mathcal{T}}
\nc{\cQ}{\mathcal{Q}}
\nc{\cW}{\mathcal{W}}
\nc{\cR}{\mathcal{R}}
\nc{\cC}{\mathcal{C}}

\nc{\pt}{\mathcal{P}\mathcal{T}}


\nc{\C}{\mathbb{C}}
\nc{\Q}{\mathbb{Q}}
\nc{\R}{\mathbb{R}}
\nc{\Z}{\mathbb{Z}}
\nc{\N}{\mathbb{N}}


\nc{\fg}{\mathfrak{g}}


\nc{\barx}{\bar{x}}
\nc{\bi}{\bar{i}}
\nc{\bj}{\bar{j}}
\nc{\bgr}{\bar{\rho}}
\nc{\bA}{\bar{\alpha}}
\nc{\bB}{\bar{\beta}}
\nc{\bC}{\bar{\gamma}}
\nc{\by}{\bar{y}}


\nc{\tf}{\tilde{f}}
\nc{\te}{\tilde{e}}
\nc{\ts}{\tilde{s}}
\nc{\tgP}{\widetilde{\Phi}}
\nc{\tgPs}{\tilde{\Psi}}
\nc{\tgn}{\tilde{\nu}}
\nc{\tgl}{\tilde{\lambda}}
\nc{\tge}{\tilde{\eta}}
\nc{\txi}{\tilde{\xi}}
\nc{\tep}{\tilde{\epsilon}}


\nc{\cB}{\check{b}}

\thispagestyle{empty}
\rightline{\small EMPG-07-06}
\vspace*{2cm}
\begin{center}
{ \LARGE Complex WKB Analysis of a $\pt$ Symmetric Eigenvalue Problem}\\
\vspace*{1cm}
\renewcommand{\thefootnote}{\fnsymbol{footnote}}
{\bf Mark~Sorrell}\footnote[1]{E-mail: {\tt
sorrell@ma.hw.ac.uk}}\\
\renewcommand{\thefootnote}{\arabic{footnote}}
\vspace*{0.3cm} \vspace{0.5cm}
{\it Department of Mathematics and the Maxwell Institute for Mathematical Sciences \\
Heriot-Watt University, Edinburgh,  EH14 4AS, UK}\\
\vspace{1.5cm}
{\large March 2007}\\
\vspace{2cm}
{\bf Abstract}
\end{center}
{The spectra of a particular class of $\pt$ symmetric eigenvalue problems has previously been studied, and found to have an extremely rich structure.  In this paper we present an explanation for these spectral properties in terms of quantisation conditions obtained from the complex WKB method.  In particular, we consider the relation of the quantisation conditions to the reality and positivity properties of the eigenvalues.  The methods are also used to examine further the pattern of eigenvalue degeneracies observed by Dorey {\it et al.} in \cite{MR1857169,Dorey:2001hi}}.
\setcounter{page}{0}
\setcounter{footnote}{0}
\newpage
\section{Introduction}\label{intro}
\begin{figure}[here!]
\begin{center}
\resizebox{5cm}{!}{\includegraphics{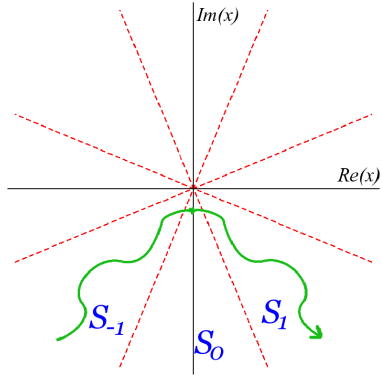}}
\caption{{\bf The quantisation contour, $\mathcal{C}$ ($M=3$)} {\em The Stokes structure is shown in the limit of large $|x|$, without showing the detailed structure near the turning points.}\label{contour}}
\end{center}
\end{figure}

\begin{figure}[here!]
\begin{center}
\resizebox{6.9cm}{!}{\includegraphics{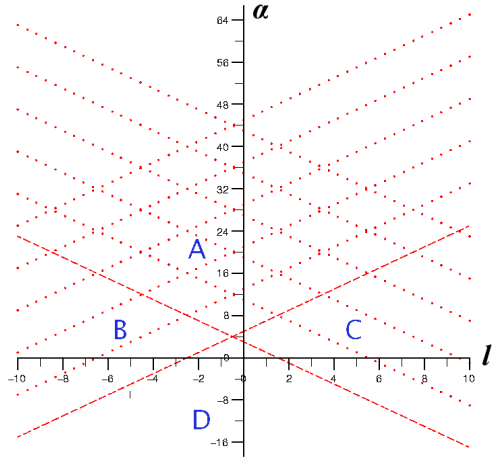}}
\caption{{\bf Regions of the $\ga$ - $l$ plane}.\hspace{0.5cm}{\em In} \cite{MR1857169}, {\em Dorey et al proved the reality of the spectrum for $(\ga,l) \in B \cup C \cup D$, and positivity for $(\ga,l) \in D$.  The dotted lines show $\ga_{\pm} \in \Z$}\label{domain0}.}
\end{center}
\end{figure}

\begin{figure}[here!]
\begin{center}
\resizebox{6cm}{!}{\includegraphics{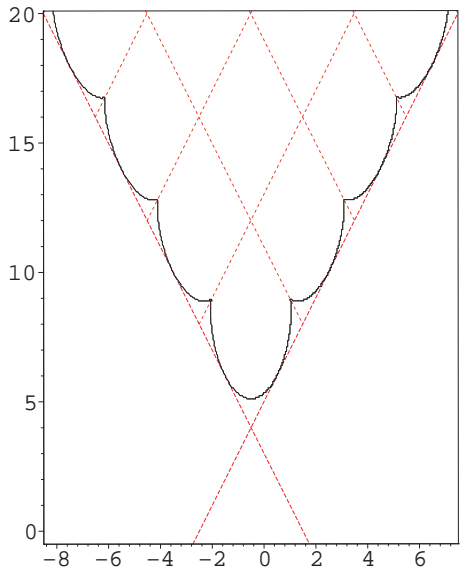}}
\caption{{\bf Domain of unreality}.\hspace{0.5cm}{\em The outline of the region where complex eigenvalues first appear, as shown in} \cite{MR1857169,Dorey:2001hi}.\label{scan}}
\end{center}
\end{figure}

The one-dimensional, time-independent Schr\"{o}dinger equation\\
\bea
-\ddx{\Psi(x)}+V(x)\Psi(x) &=& E\Psi(x),  \nn \\ \nn
\eea
with  $\Psi(x) \in L^2(\cC)$ ({\it i.e.} $\Psi(x)$ is square integrable over some given contour $\cC$ in the complex plane), is an eigenvalue problem for the energy $E$. Eigenvalue problems of this type are a natural extension of eigenvalue problems on the real line \cite{Bender:1992bk}, and indeed, the usual requirement that  $\Psi(x) \in L^2(\cR)$ just corresponds to a particular choice of contour. \\
In recent years, there has been considerable interest in the study of such eigenvalue problems associated with Hamiltonians that have the property of $\mathcal{PT}$ symmetry \cite{MR1627442, MR1686605, MR1742959}.  The $\mathcal{PT}$ operator is an anti-linear operator corresponding to parity reflection and time reversal: \\
\bea
\mathcal{P} \Phi(x)=\Phi(-x) \nn \\[0.2cm]
\mathcal{T} \Phi(x)=\Phi^*(x) \nn \\ \nn
\eea
A reason for the interest in $\pt$ symmetry is that the spectrum of $\mathcal{PT}$ symmetric Hamiltonians sometimes appears to be purely real.  In general,  it can be shown that the eigenvalues of these Hamiltonians will be real, or occur in complex conjugate pairs.  A connection has also been recognised between these types of differential equations and the field of integrable models (the `ODE/IM correspondence')  \cite{Dorey:1998pt,MR1832065,Dorey:1999uk,MR1832082,Dorey:2000kq,Dorey:2004ta}. \\ \\ \\ \\
In this paper we will consider a particular $\pt$ symmetric eigenvalue problem\\
\bea \label{eig}
\left[ -\frac{d^2}{dx^2}+x^6+\ga x^2+\frac{l(l+1)}{x^2} \right] \Psi(x) &=& \gl \Psi(x),\hspace{1cm} \Psi(x) \in L^2(\mathcal{C}) \\ \nn
\eea
This is the $M=3$ case of the family of equations\\
\bea
\left[ -\frac{d^2}{dx^2}-(ix)^{2M}-\ga (ix)^{M-1}+\frac{l(l+1)}{x^2} \right] \Psi(x) &=& \gl \Psi(x),\hspace{1cm} \Psi(x) \in L^2(\mathcal{C}) \nn \\[0.2cm] \nn
\eea
discussed in \cite{MR1857169,Dorey:2001hi}, with the contour $\cC$ joining $|x|=\infty$ in the Stokes sectors  $\cS_{-1}$ and $\cS_{1}$ (Figure \ref{contour}). \\[2.8cm]
\\
It was shown in \cite{MR1857169,Dorey:2001hi} that this equation, with these boundary conditions, has a spectrum which is:
\begin{center}
\hspace{0.8cm} {\it real} \hspace{0.2cm} if $\ga <M+1+|2l+1|$  \hspace{1cm} [Regions $B,C,D$]\\
{\it positive} if $\ga<M+1-|2l+1|$ \hspace{1cm} [Region D] .
\end{center}
However, for  $\ga >M+1+|2l+1|$, (Region $A$), complex energy levels may be found.  For the $M=3$ case, these four regions in the $(\ga,l)$ plane are shown in Figure \ref{domain0}, separated by the dashed lines.
\\
Also shown in \cite{MR1857169,Dorey:2001hi} was an interesting structure to the boundary of the region, within $A$, where eigenvalues become complex (the `domain of unreality'), notably the appearance of `cusps' on this boundary (Figure \ref{scan}).\\ \\
\\
The remainder of this paper will be structured as follows: Section \ref{wkb} deals with the complex WKB method, and how quantisation conditions derived using this method are able to account for the results found in \cite{MR1857169,Dorey:2001hi}.  In particular, in \ref{wkb1}, we give a brief summary of the complex WKB method and its use.  Section \ref{qc} explains how this method can be used to obtain quantisation conditions for the energy eigenvalues, and in \ref{qcprob} we describe how different quantisation conditions are applicable in this problem for different values of the parameters $E,\ga$ and $l$.  Section \ref{examplesec} gives an example of calculating a quantisation condition for some particular values of the parameters. In section \ref{pos} we demonstrate that the WKB quantisation condition approach is able to explain the positivity of the spectrum in the region found in \cite{MR1857169,Dorey:2001hi}.  Section \ref{rea} explains the appearance of complex eigenvalues in terms of the quantisation conditions, in the region where Dorey {\it et al.} found they could occur, \cite{MR1857169,Dorey:2001hi} , and we conjecture how complex WKB methods might demonstrate reality.  In section \ref{deg} we use the WKB quantisation conditions to calculate the positions of degenerate eigenvalues (and reproduce the boundary of the domain of unreality), and explain the formation of the cusps.  Our results are compared to those found in \cite{MR1857169,Dorey:2001hi}.  Finally, Section \ref{conc} contains conclusions and a discussion of possible future work. \\ \\
As a brief aside, in \cite{Dorey:2001hi}, Dorey {\it et al.} defined a new set of coordinates on the $(\ga,l)$ plane by
\bea
\ga_{\pm}=\frac{1}{2M+2}[\ga -M-1 \pm (2l+1)] \nn
\eea 
which we will also occasionally adopt in order to allow an easier comparison of results with \cite{Dorey:2001hi}.  In particular, for the M=3 case dealt with here,
\bea
\ga_{\pm}=\frac{1}{8}[\ga -4 \pm (2l+1)] \nn
\eea 
so that 
\bea
\ga=4(1+\ga_{+}+\ga_{-}) \mbox{ and } l=\frac{1}{2}[4(\ga_{+}-\ga_{-})-1] \nn
\eea
The dotted lines shown in Figures \ref{domain0} and \ref{scan} indicate $\ga_{\pm} \in \Z$.
\section{Complex WKB}\label{wkb}
\subsection{General Principles}\label{wkb1}
For the one-dimensional, time-independent Schr\"{o}dinger equation
\bea
-\ddx{\Psi(x)}+V(x)\Psi(x) &=& E\Psi(x),  \nn 
\eea
the  first-order WKB approximation (see {\it e.g.} \cite{Griffiths}) states that two solutions to this equation are given asymptotically for $|x| \to \infty $ by 
\begin{equation}
\label{wkbapprox}
\Psi_{\pm}(x) \sim \frac{1}{P(x)^{\frac{1}{4}}}\exp\left(\pm i \int_{x_0}^x(P(t))^{\frac{1}{2}}dt\right),\hspace{1cm} P(x)=E-V(x).
\end{equation}
The complex WKB method (see {\it e.g.} \cite{Heading:1962,Berry:1972na,Voros83}) involves treating $x$ as a complex variable, and calculating how the asymptotic form of the solutions change as they are traced around the complex plane. The lower bound of the integral $x_0$ will be a zero of $P(x)$ ({\it i.e.} a turning point of the classical problem), and the solutions are valid in some region of the complex plane around $x_0$.\\    
The integral in the exponent is, in general, complex valued.  Because it will typically have an imaginary part, the solutions will contain real exponentials.  If a solution contains a growing exponential term, it is said to be dominant, while a solution with a decaying exponential is called subdominant.  However, if this integral is purely real, then both WKB solutions will be purely oscillatory, with neither dominating the other.  Lines can be drawn in the complex plane, emanating from the turning points (zeros),  marking the curves where ${\Im m}\left[\int_{x_0}^x(P(t))^{\frac{1}{2}}dt)\right]=0$.  These are known as {\it anti-Stokes} lines.  Anti-Stokes lines mark the borders between sectors of dominant/subdominant behaviour;  in a given sector one solution will be dominant, but on crossing an anti-Stokes line to the next sector this behaviour reverses and the solution will be subdominant in the new sector.\\
We can also find the regions where ${\Re e}\left[\int_{x_0}^x(P(t))^{\frac{1}{2}}dt)\right]=0$.  These are known as {\it Stokes} lines, and are the regions where the solutions are most dominant/subdominant.  Knowing the position of these Stokes and anti-Stokes lines is crucial in applying complex WKB methods. \\ \\
Beginning with a linear combination of the two solutions at one point of the complex plane, a globally defined solution may be obtained by following several well-established rules.  The most important of these for our purposes can be summarised:\\
\begin{enumerate}
\item When crossing an anti-Stokes line, the solutions `exchange dominance', {\it i.e.} a dominant solution becomes subdominant, and vice versa.\\
\item Upon crossing a Stokes line, the coefficient of the subdominant solution changes by  an amount proportional to the coefficient of the dominant term.
\begin{center}
{\it i.e.} {\bf new sub. coefficient $=$ old sub. coefficient $+$ $T$  $\times$ old dom. coefficient}\\[0.4cm]
\end{center}
$T$ is called a Stokes multiplier, and depends on the nature of the turning point that the Stokes line originates from.  For a Stokes line emanating from a simple turning point the value of the Stokes multiplier $T$ is $i$, for a Stokes line from a double zero, $T=\sqrt{2}i$ (for the first-order approximation). 
The coefficient of the dominant term remains unchanged when a Stokes line is crossed.\\
\item The rules given in 1. and 2. refer to a WKB solution defined in terms of a particular turning point crossing a Stokes/anti-Stokes line emanating from that turning point. If it is intended to continue a solution across a line from a different turning point, the WKB solution must first be rewritten in terms of this new zero.  To connect solutions defined in terms of different turning points, $x_1$ and $x_2$, use:\\
\begin{center}
$ \exp\left(\pm i \int_{x_1}^x(P(t))^{\frac{1}{2}}dt\right) = \exp\left(\pm i \int_{x_1}^{x_2}(P(t))^{\frac{1}{2}}dt\right) \exp\left(\pm i \int_{x_2}^x(P(t))^{\frac{1}{2}}dt\right).$\\
\end{center}
\end{enumerate}
Note that it is chosen to introduce the discontinuous change in the subdominant coefficient at the Stokes line, so that this discontinuity is small compared to the error in the WKB approximation. \\
\subsection{WKB Quantisation conditions}\label{qc}
We study the spectrum of this eigenvalue problem via quantisation conditions obtained using the complex WKB method.  Briefly, this is done by starting with a WKB solution that is purely subdominant in the Stokes sector $\cS_{-1}$ (Figure \ref{contour}).  This solution can then be traced around the complex plane, in an anti-clockwise direction, with new subdominant exponentials appearing as Stokes lines are crossed.  The new subdominant contributions will have a coefficient of $i$ (the Stokes multiplier associated with line from a simple turning point) or $\sqrt{2}i$ (the Stokes multiplier associated with line from a second order zero) multiplied by the coefficient of the dominant exponential at that Stokes line (in accordance with the established complex WKB rules - see for instance \cite{Heading:1962, Berry:1972na}).  This leads to a wave function in the Stokes sector $\cS_{1}$ that has both a subdominant and dominant component.  Requiring that the coefficient of the dominant solution vanishes (to fulfil the boundary conditions) leads to a quantisation condition for the energy.  This is an approach used in \cite{BenderBerry:2001wk} to investigate the appearance of complex conjugate eigenvalues in another $\pt$ symmetric problem.  An explicit example of this type of calculation is shown in Section \ref{examplesec}.\\
\subsection{WKB Quantisation conditions for this eigenvalue problem}\label{qcprob}
\begin{figure}[h!]
\begin{center}
\resizebox{17cm}{!}{\includegraphics{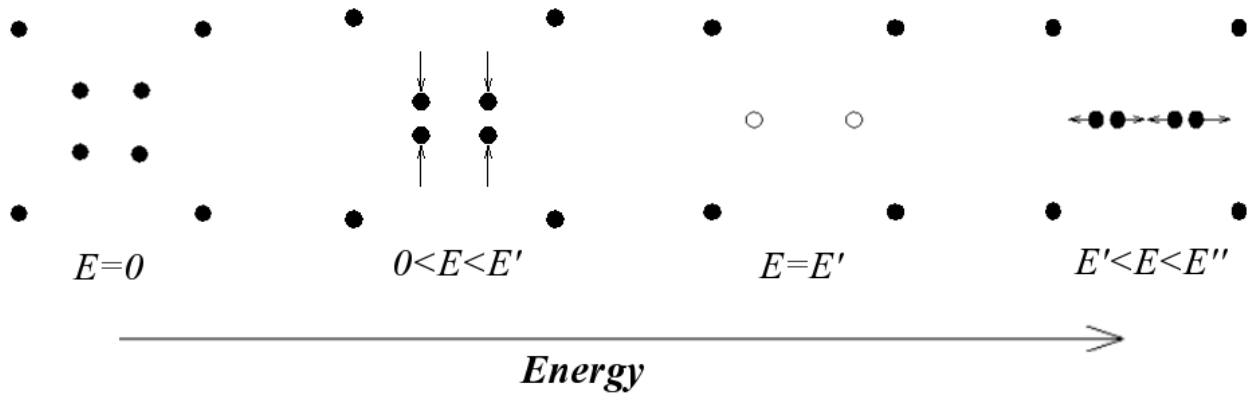}}
\end{center}
\caption{{\bf Changes in turning point configuration occurring with increasing energy.}  {\em Example shown is for $\ga =3, l=0.5$.  This corresponds to the first row in Table \ref{plus}.  $E^\prime$ is the energy value at which the zeros coalesce to form a double zero.  $E^{\prime\prime}$ is the energy value at which the outer of the zeros in the horizontal line `line up' vertically with the outer zeros.  Both $E^\prime$ and $E^{\prime\prime}$ can be expressed as a function of $\ga$ and $l$.  For these values of $\ga$ and $l$ we have $E^\prime \approx 3.1075, E^{\prime\prime} \approx3.5905$.}  \label{config}}.\hspace{0.5cm}
\end{figure}

\begin{figure}[h!]
\begin{center}
\resizebox{12cm}{!}{\includegraphics{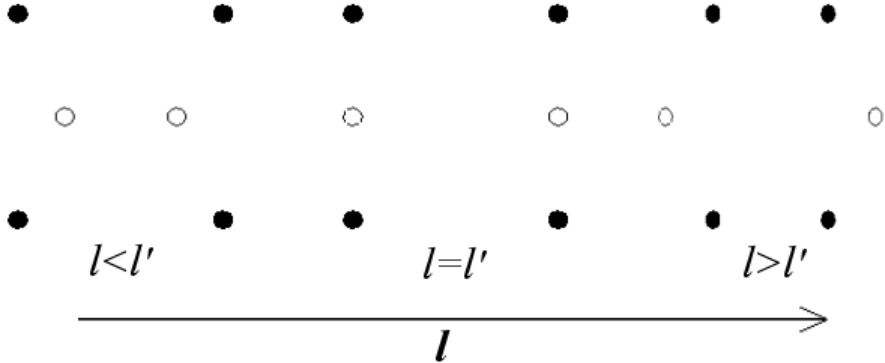}}
\caption{{\bf Relative positions of double zeros.}  {\em  $\ga >0, E=E^\prime$.  This corresponds to the $E=E^\prime$ column in Table \ref{plus}. }  \label{configl}}\hspace{0.5cm}
\end{center}
\end{figure}

We now specialise our discussion to the eigenvalue problem of eq. (\ref{eig}).  Complications arise in this problem because many different arrangements of turning points, and hence different topologies of Stokes and anti-Stokes lines, are found for different values of the parameters in the equation.  This leads to different quantisation conditions being required for different ranges of values of $\ga$, $l$ and $E$.  These different possible arrangements are summarised in Table \ref{plus} and Table \ref{minus}. Table \ref{plus} refers to positive values of $\ga$, and Table \ref{minus} to negative.  Each table is split into five rows, by different ranges of values of $l$, relative to the key values $l=0$ and $l=l^\prime$.  The remainder of this section will now be used to explain the origin of this $l^\prime$, and the way the different Stokes structures are organised according to the parameter values.\\ \\
As energy is increased (or decreased) from $0$, pairs of turning points move together, until an energy value is reached at which a double zero is found.  The double zero then splits, and the two turning points move off again, perpendicular to their original path. (See Figure \ref{config}).\\[0.8 cm]
\\
Different Stokes structures are obtained according to where this coalescence occurs in respect to the other turning points.  {\it i.e.} new topologies are obtained depending on whether the double zeros occur within, outside of, or directly in line with the other turning points.  This can be easily seen by looking at Figure \ref{configl} (which corresponds to looking down the $E=E^\prime$ column of Table \ref{plus}); In the first entry the double zeros appear inside the other four zeros, in the second they line up vertically with the other zeros and in the third the double zeros are outside the box formed by the remaining zeros.  \\
\\
For a given value of the parameter $\ga$, the value of $l$ required for the zeros that coalesce to line up with the others, (which we will denote $l^\prime$), is given by
\bea \label{l}
l^\prime= -\frac{1}{2}+\frac{\sqrt{(1+\ga ^2)}}{2}
\eea
If $l$ is less than this $l^\prime$, turning points move together within the remaining turning points (in a horizontal sense for positive energies and vertical sense for negative energies).  For $l>l^\prime$, the turning points that coalesce are positioned outside of the others (again in a horizontal sense for $E>0$ and vertical for $E<0$).  \\ 
Another important value of $l$ where the behaviour changes is $l=0$. For $l=0$ and $l<0$, different sequences of arrangements occur, and these are shown in the second and third blocks of Tables \ref{plus} and \ref{minus}.  Obviously for $l=0$ (and $l=-1$), the order $x^{-2}$ term vanishes from the potential, and there are now only six zeros, instead of eight. \\ 
\begin{landscape}
\thispagestyle{empty}
\renewcommand{\thefootnote}{\fnsymbol{footnote}}
\begin{table}
\small
\caption{\bf Arrangements of Turning Points, $\ga>0$ \label{plus}}
\begin{center} 
\begin{tabular}{|c|c|c|c|c|c|c|c|c|c|c|c|c|}
\hline \hline 
\scriptsize Region & {\bf } & \scriptsize {\bf $E<-E^{\prime\prime}$} & \scriptsize {\bf $E=-E^{\prime\prime}$} & \scriptsize {\bf $-E^{\prime\prime}<E<-E^\prime$} &  \scriptsize {\bf $E=-E^\prime$} & \scriptsize {\bf $-E^\prime<E<0$} & \scriptsize {\bf $E=0$} & \scriptsize {\bf $0<E<E^\prime$} & \scriptsize {\bf $E=E^\prime$} & \scriptsize {\bf $E^\prime<E<E^{\prime\prime}$} & \scriptsize {\bf $E=E^{\prime\prime}$} & \scriptsize {\bf $E>E^{\prime\prime}$} \footnotemark[1] \\  
\scriptsize (Figure \ref{domain1}) & & & & & & & & & & & & \\ \hline
\hline
{\bf $I$} & {\scriptsize \bf $0<l<l^\prime$} & \resizebox{0.7cm}{!}{\includegraphics{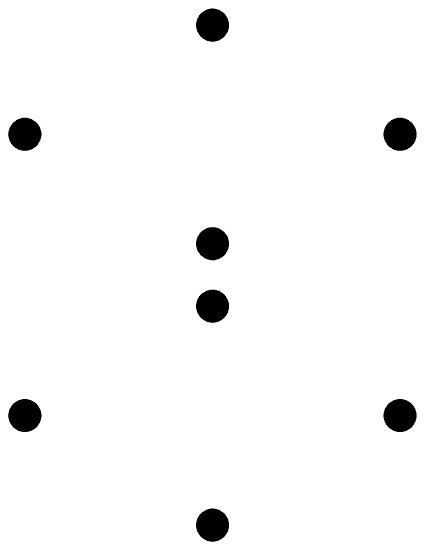}} & \resizebox{0.8cm}{!}{\includegraphics{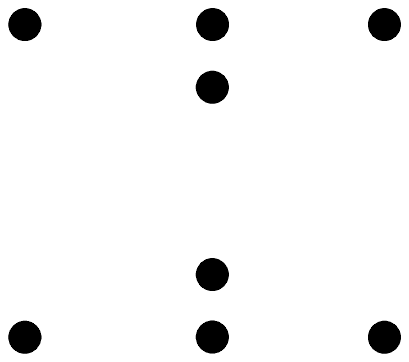}} & \resizebox{0.7cm}{!}{\includegraphics{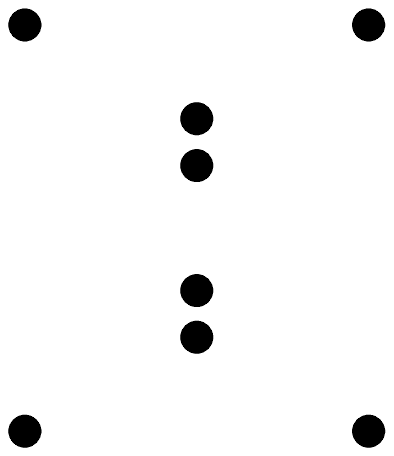}} & \resizebox{0.8cm}{!}{\includegraphics{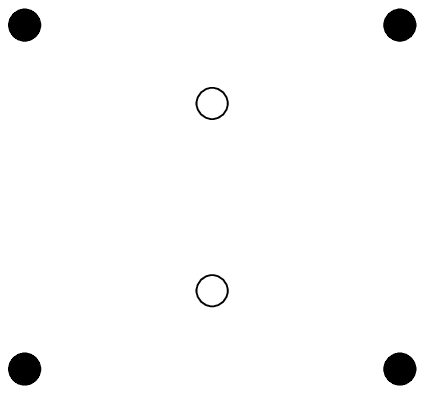}} & \resizebox{0.8cm}{!}{\includegraphics{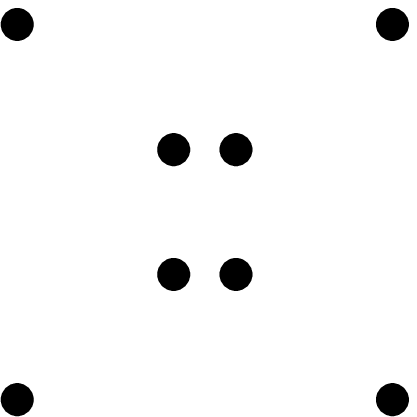}} &\resizebox{0.8cm}{!}{\includegraphics{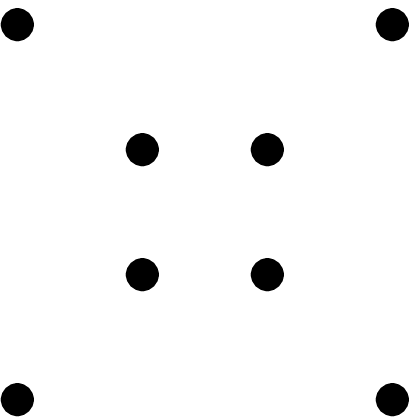}} & \resizebox{0.8cm}{!}{\includegraphics{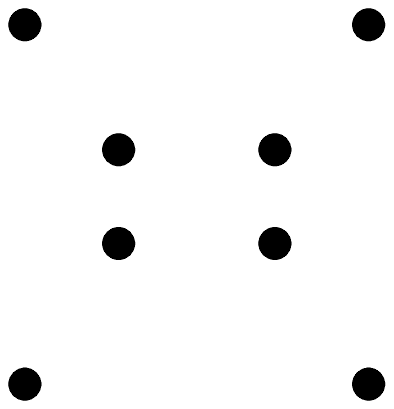}} & \resizebox{0.8cm}{!}{\includegraphics{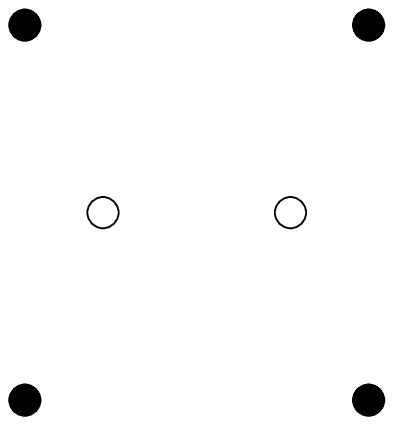}} &\resizebox{0.9cm}{!}{\includegraphics{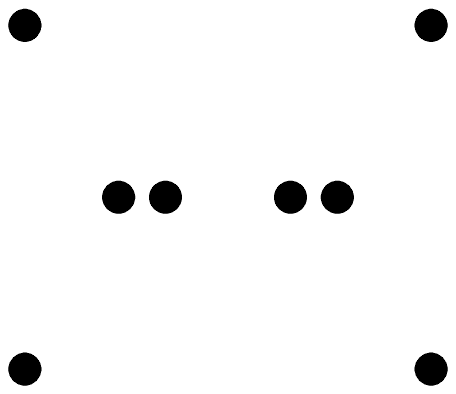}} &\resizebox{0.7cm}{!}{\includegraphics{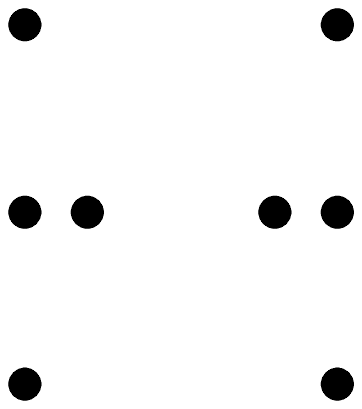}} & \resizebox{0.8cm}{!}{\includegraphics{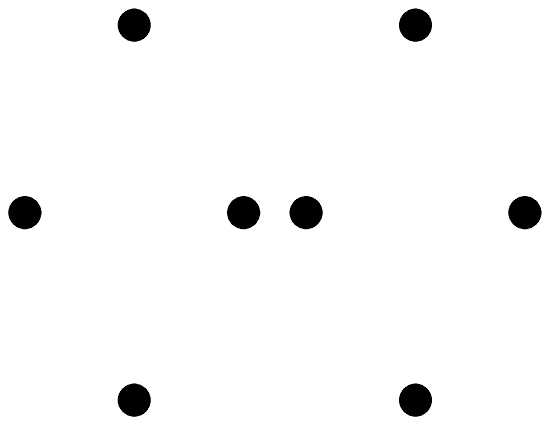}} \\ \hline
\scriptsize {\bf Boundary of $H/I$} & \scriptsize {\bf $l=l^\prime$} & \resizebox{0.7cm}{!}{\includegraphics{top3}}  & \footnotemark[2] &  \footnotemark[2] & \resizebox{0.7cm}{!}{\includegraphics{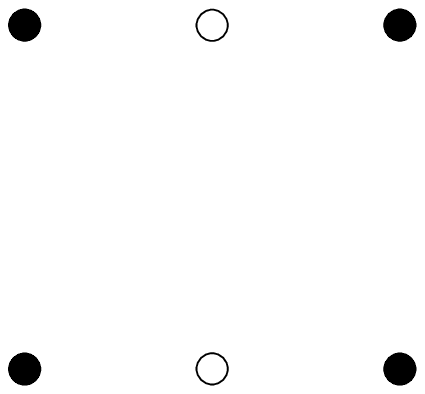}} & \resizebox{0.7cm}{!}{\includegraphics{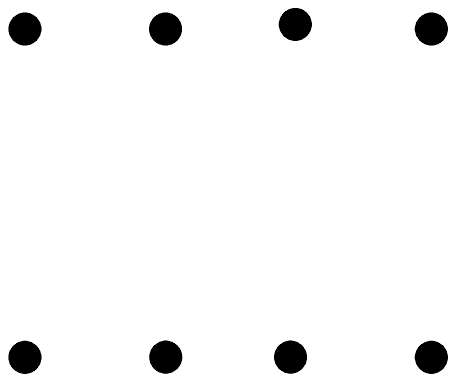}} & \resizebox{0.7cm}{!}{\includegraphics{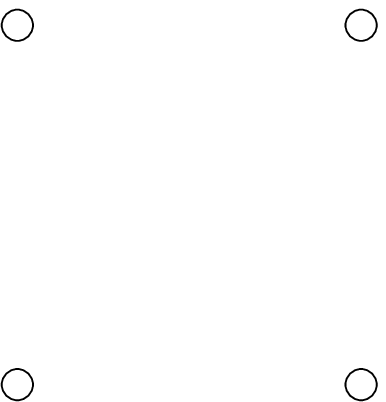}} &  \resizebox{0.7cm}{!}{\includegraphics{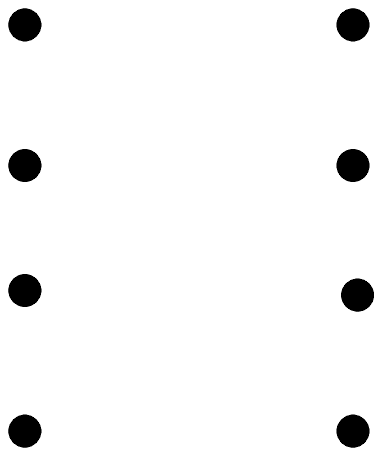}} & \resizebox{0.7cm}{!}{\includegraphics{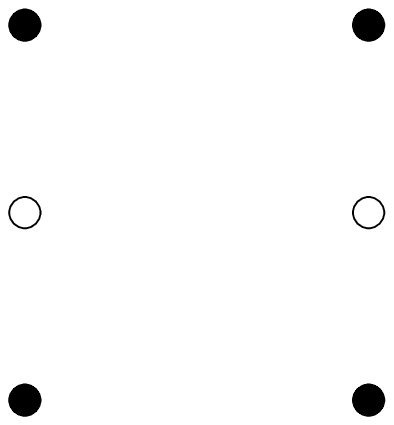}} & \footnotemark[8] & \footnotemark[8]  & \resizebox{0.8cm}{!}{\includegraphics{top11}} \\ \hline
{\bf $H$} & \scriptsize {\bf $l>l^\prime$}& \resizebox{0.7cm}{!}{\includegraphics{top3}} & \resizebox{0.7cm}{!}{\includegraphics{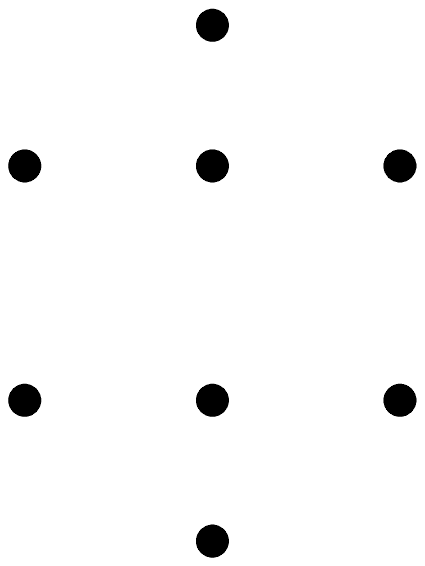}} & \resizebox{0.7cm}{!}{\includegraphics{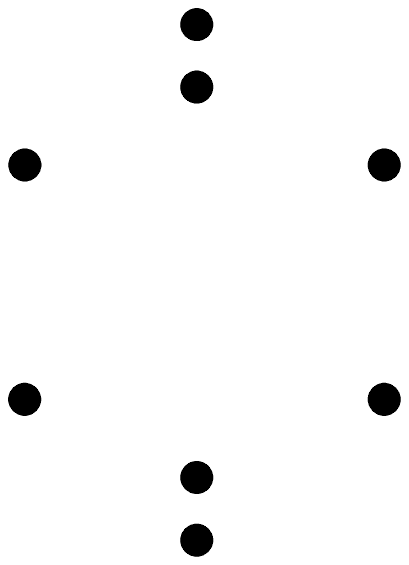}} & \resizebox{0.7cm}{!}{\includegraphics{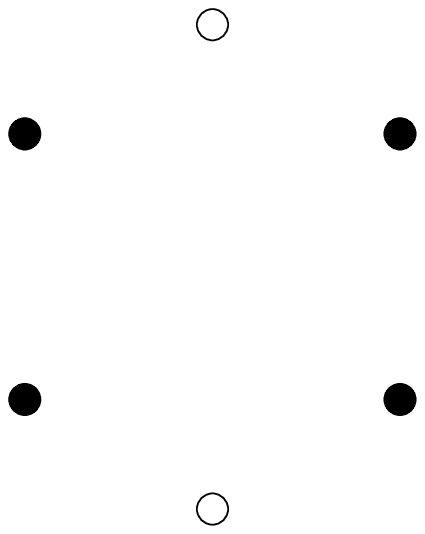}} & \resizebox{0.8cm}{!}{\includegraphics{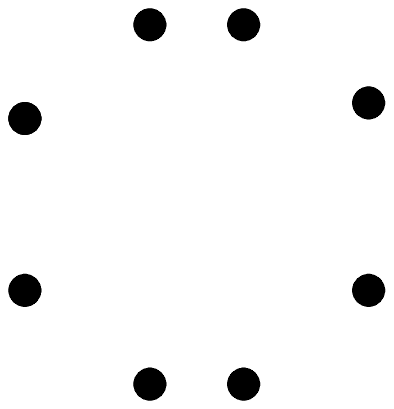}} & \resizebox{0.8cm}{!}{\includegraphics{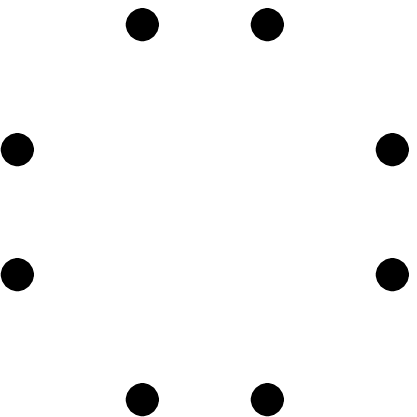}} & \resizebox{0.8cm}{!}{\includegraphics{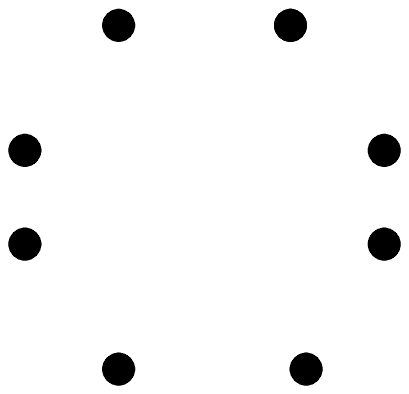}} & \resizebox{0.8cm}{!}{\includegraphics{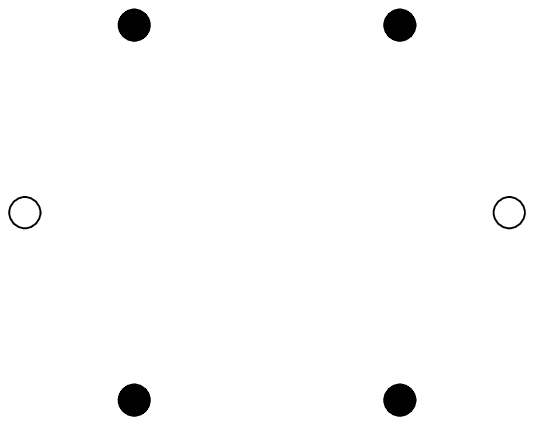}} & \resizebox{0.8cm}{!}{\includegraphics{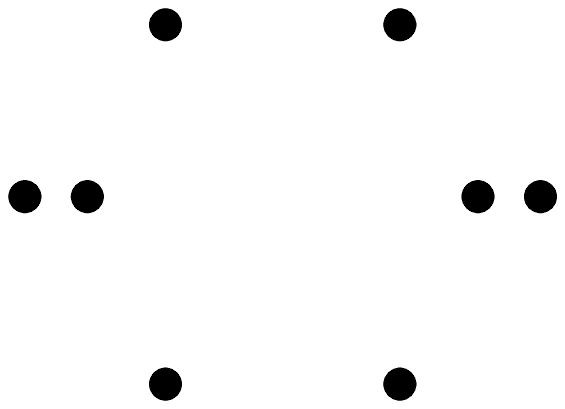}} & \resizebox{0.8cm}{!}{\includegraphics{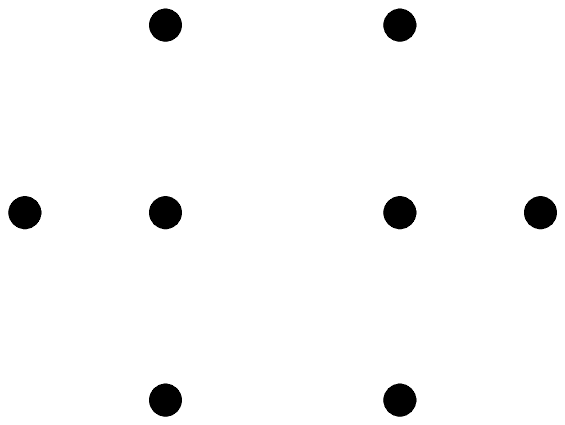}} & \resizebox{0.8cm}{!}{\includegraphics{top11}} \\ \hline
\end{tabular}
\begin{tabular}{ccc}
 \\
\end{tabular}

\begin{tabular}{|c|c|c|c|c|c|c|c|c|}
\hline
\scriptsize Region & {\bf } & \scriptsize {\bf $E<-E^{\prime\prime\prime}$} & \scriptsize {\bf $E=-E^{\prime\prime\prime}$} & \scriptsize {\bf $-E^{\prime\prime\prime}<E<0$} & \scriptsize {\bf $E=0$} & \scriptsize {\bf $0<E<E^{\prime\prime\prime}$} & \scriptsize {\bf $E=E^{\prime\prime\prime}$} & \scriptsize {\bf $E>E^{\prime\prime\prime}$} \\ 
\scriptsize (Figure \ref{domain1}) &  & & & & & & &  \\ \hline \hline
{\bf $J$} & \scriptsize {\bf $-\frac{1}{2}<l<0$} & \resizebox{0.7cm}{!}{\includegraphics{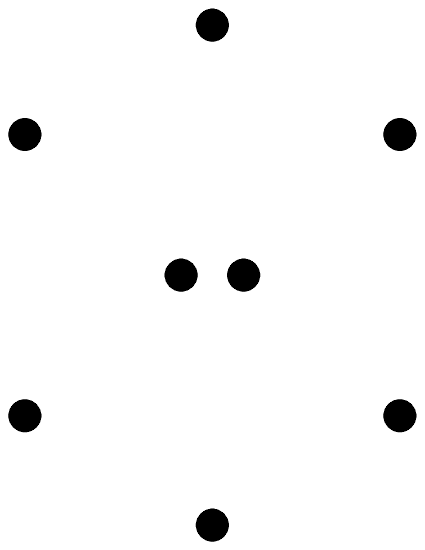}} & \resizebox{0.8cm}{!}{\includegraphics{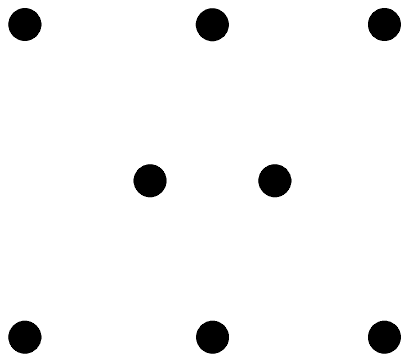}} & \resizebox{0.7cm}{!}{\includegraphics{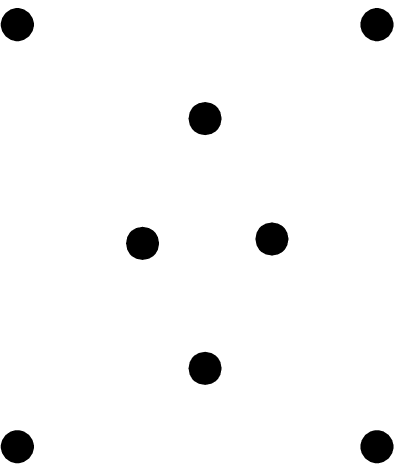}} & \resizebox{0.8cm}{!}{\includegraphics{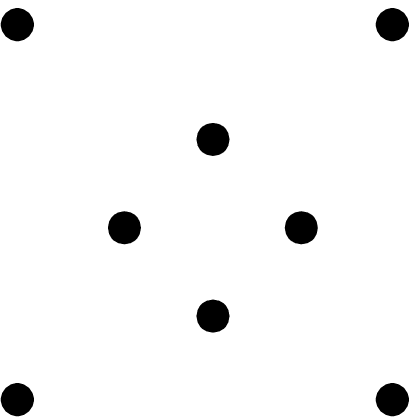}} & \resizebox{0.8cm}{!}{\includegraphics{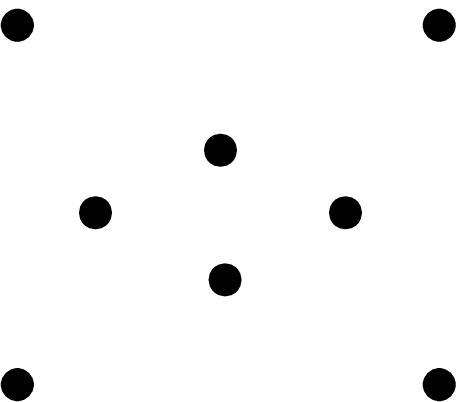}} & \resizebox{0.7cm}{!}{\includegraphics{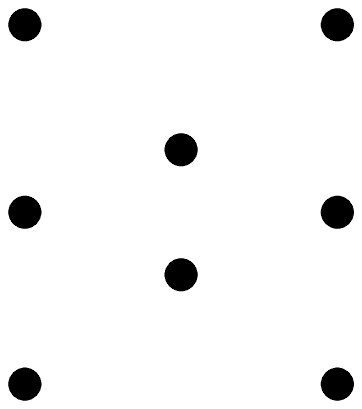}} & \resizebox{0.8cm}{!}{\includegraphics{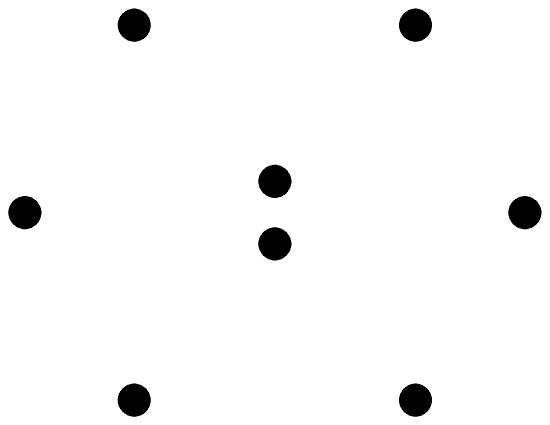}}
 \\ \hline
\end{tabular}

\begin{tabular}{ccc}
 \\
\end{tabular}

\begin{tabular}{|c|c|c|c|c|c|c|c|c|c|}
\hline
\scriptsize Region & & \scriptsize {\bf $E<-E^{\prime\prime}$} & \scriptsize {\bf $E=-E^{\prime\prime}$} & \scriptsize {\bf $-E^{\prime\prime}<E<0$} & \scriptsize {\bf $E=0$} & \scriptsize {\bf $0<E<E^\prime\prime$} & \scriptsize {\bf $E=E^{\prime\prime}$} & \scriptsize {\bf $E>E^{\prime\prime}$} \footnotemark[1] \\ 
 \scriptsize (Figure \ref{domain1}) & & & & & & & & \\ \hline \hline
 \scriptsize {\bf Boundary of $I/J$} &  \scriptsize {\bf $l=0$} & \resizebox{0.7cm}{!}{\includegraphics{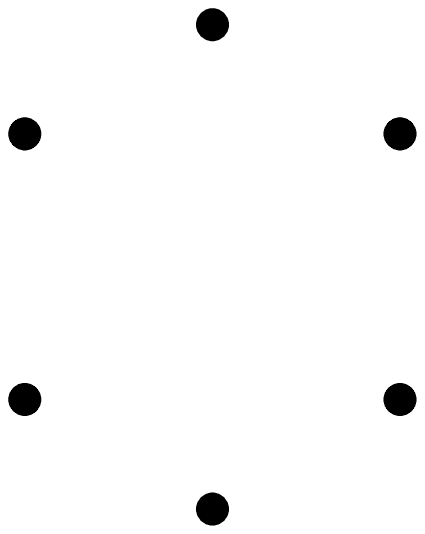}} & \resizebox{0.8cm}{!}{\includegraphics{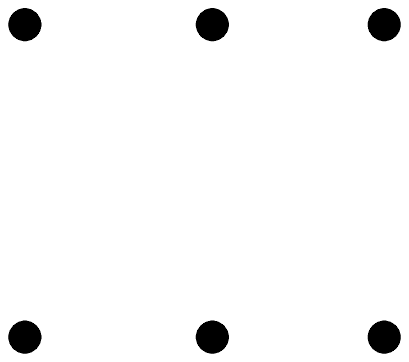}} & \resizebox{0.8cm}{!}{\includegraphics{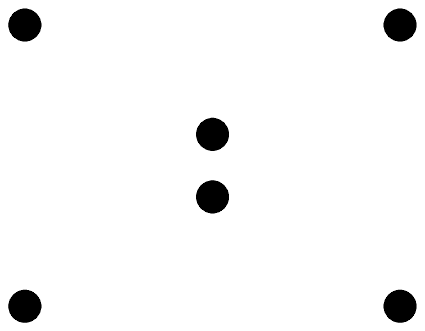}} & \resizebox{0.7cm}{!}{\includegraphics{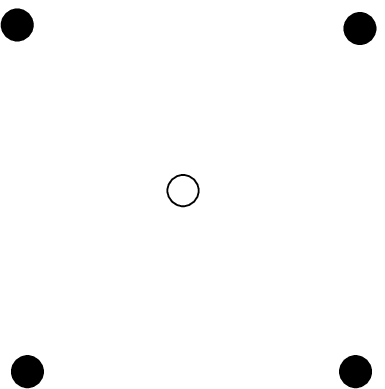}} & \resizebox{0.7cm}{!}{\includegraphics{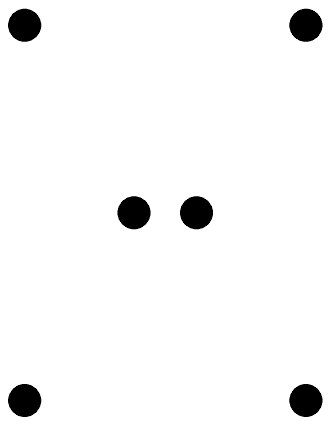}} & \resizebox{0.7cm}{!}{\includegraphics{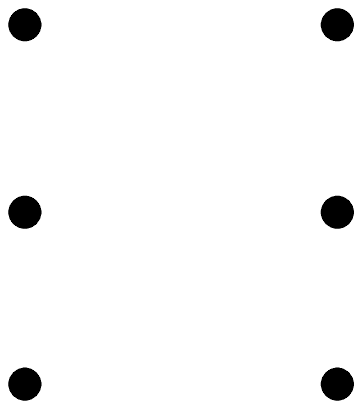}} & \resizebox{0.8cm}{!}{\includegraphics{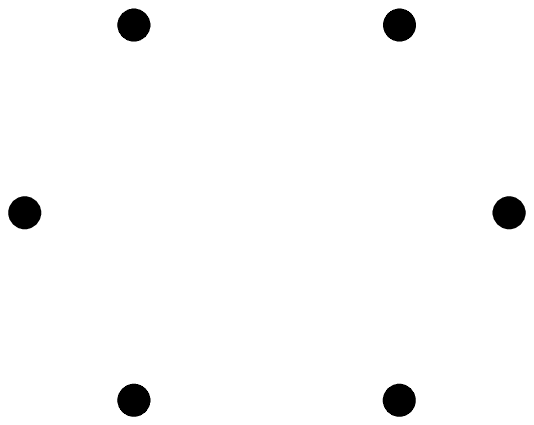}} \\ \hline
\end{tabular}
\end{center}
\end{table}
\footnotetext[2]{$-E^\prime=-E^{\prime\prime}$ for $l=l^\prime$}
\footnotetext[8]{$E^\prime=E^{\prime\prime}$ for $l=l^\prime$}
\footnotetext[1]{Change in Stokes structure actually occurs at a value of $E$ slightly larger than $E^{\prime\prime}$.  Hence the quantisation condition associated with this arrangement does not come into effect exactly at $E=E^{\prime\prime}$}
\end{landscape}
\renewcommand{\thefootnote}{\arabic{footnote}}
\begin{landscape}
\thispagestyle{empty}
\renewcommand{\thefootnote}{\fnsymbol{footnote}}
\begin{table}
\small
\caption{\bf Arrangements of Turning Points, $\ga<0$ \label{minus}}
\begin{center} 
\begin{tabular}{|c|c|c|c|c|c|c|c|c|c|c|c|c|}
\hline \hline 
\scriptsize Region & {\bf } & \scriptsize {\bf $E<-E^{\prime\prime}$} & \scriptsize {\bf $E=-E^{\prime\prime}$} & \scriptsize {\bf $-E^{\prime\prime}<E<-E^\prime$} & \scriptsize {\bf $E=-E^\prime$} & \scriptsize {\bf $-E^\prime<E<0$} & \scriptsize {\bf $E=0$} & \scriptsize {\bf $0<E<E^\prime$} & \scriptsize {\bf $E=E^\prime$} & \scriptsize {\bf $E^\prime<E<E^{\prime\prime}$} & \scriptsize {\bf $E=E^{\prime\prime}$} & \scriptsize {\bf $E>E^{\prime\prime}$} \footnotemark[1]  \\
\scriptsize (Figure \ref{domain1}) & & & & & & & & & & & &\\ \hline
\hline
{\bf $E$} & \scriptsize {\bf $0<l<l^\prime$} & \resizebox{0.7cm}{!}{\includegraphics{top3}} & \resizebox{0.7cm}{!}{\includegraphics{top29}} & \resizebox{0.7cm}{!}{\includegraphics{top31}} & \resizebox{0.7cm}{!}{\includegraphics{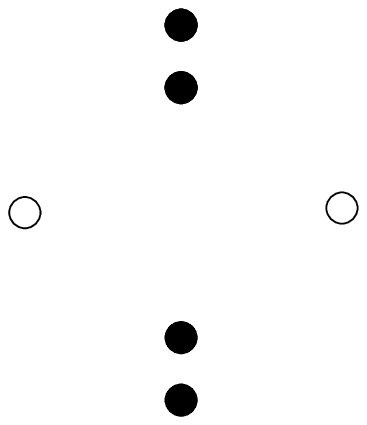}} & \resizebox{0.7cm}{!}{\includegraphics{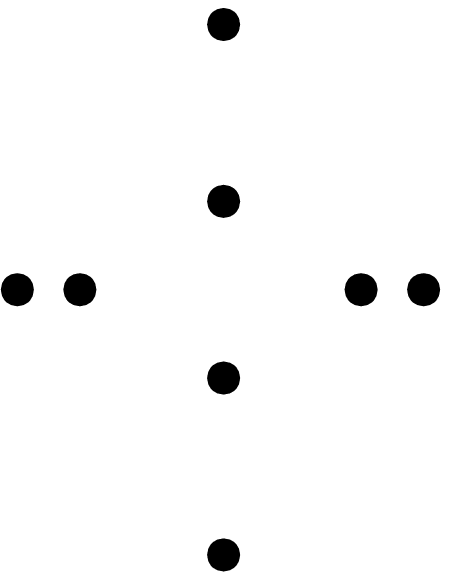}} &\resizebox{1cm}{!}{\includegraphics{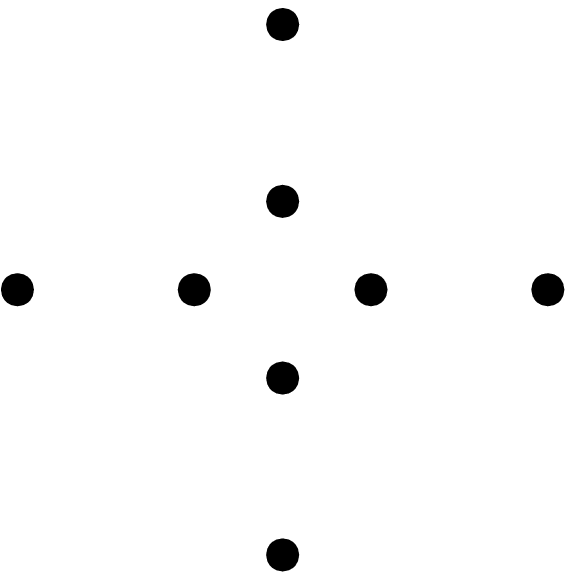}} & \resizebox{1cm}{!}{\includegraphics{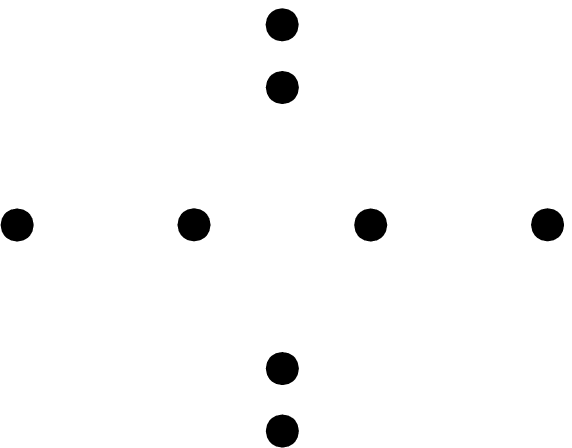}} & \resizebox{1cm}{!}{\includegraphics{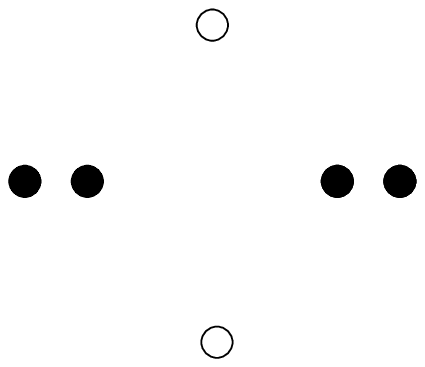}} &\resizebox{1cm}{!}{\includegraphics{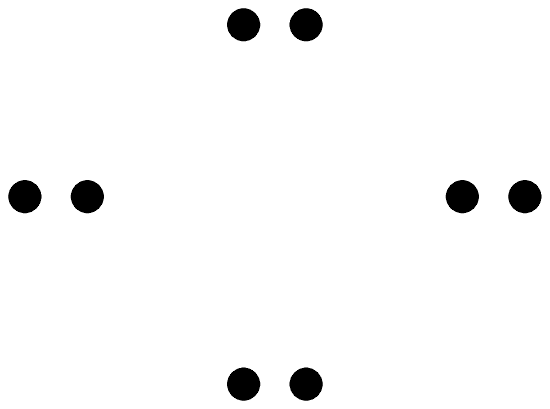}} &\resizebox{0.8cm}{!}{\includegraphics{top8}} & \resizebox{1cm}{!}{\includegraphics{top11}} \\ \hline
\tiny {\bf Boundary of $E/G$} & \scriptsize {\bf $l=l^\prime$} & \resizebox{0.7cm}{!}{\includegraphics{top3}} & \resizebox{0.7cm}{!}{\includegraphics{top29}} & \resizebox{0.7cm}{!}{\includegraphics{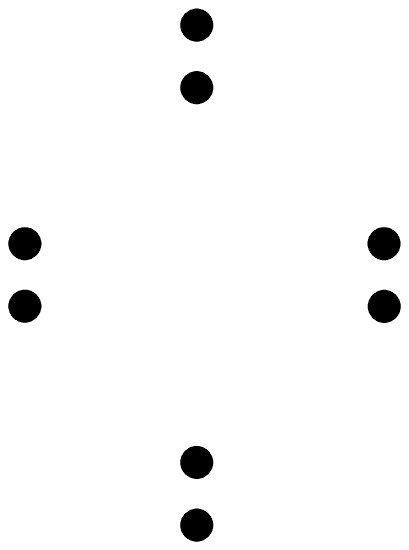}} & \footnotemark[3] &  \footnotemark[3] & \resizebox{1cm}{!}{\includegraphics{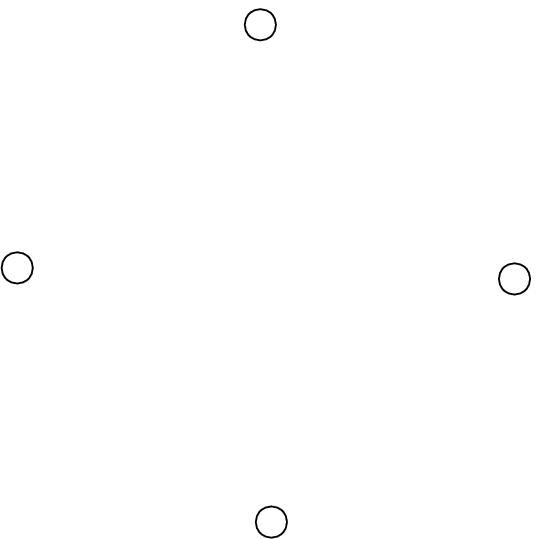}} & \footnotemark[9] & \footnotemark[9] & \resizebox{1cm}{!}{\includegraphics{top9}} & \resizebox{0.8cm}{!}{\includegraphics{top8}} & \resizebox{1cm}{!}{\includegraphics{top11}} \\ \hline
{\bf $G$ } & \scriptsize {\bf $l>l^\prime$}& \resizebox{0.7cm}{!}{\includegraphics{top3}} & \resizebox{0.7cm}{!}{\includegraphics{top29}} & \resizebox{0.7cm}{!}{\includegraphics{top31}} & \resizebox{0.7cm}{!}{\includegraphics{top30}} & \resizebox{0.7cm}{!}{\includegraphics{top20}} & \resizebox{0.7cm}{!}{\includegraphics{top32}} & \resizebox{0.7cm}{!}{\includegraphics{top2}} & \resizebox{0.7cm}{!}{\includegraphics{top6}} & \resizebox{1cm}{!}{\includegraphics{top7}} & \resizebox{0.8cm}{!}{\includegraphics{top8}} & \resizebox{1cm}{!}{\includegraphics{top11}} \\ \hline 
\end{tabular}
\begin{tabular}{ccc}
 \\ 
\end{tabular}
\begin{tabular}{|c|c|c|c|c|c|c|c|c|c|c|c|c|}
\hline
\scriptsize Region & {\bf } & \scriptsize {\bf $E<-E^{\prime}$} & \scriptsize {\bf $E=-E^{\prime}$} & \scriptsize {\bf $-E^{\prime}<E<-E_0$} & \scriptsize {\bf $E=E_0$} & \scriptsize {\bf $-E_0<E<0$} & \scriptsize {\bf $E=0$} & \scriptsize {\bf $0<E<E_0$} & \scriptsize {\bf $E=E_0$} &  \scriptsize {\bf $E_0<E<E^{\prime}$} & \scriptsize {\bf $E=E^{\prime}$} & \scriptsize {\bf $E>E^{\prime}$} \\ 
\scriptsize (Figure \ref{domain1}) & & & & & & & & & & & & \\ \hline \hline
{\bf $F$} & \scriptsize {\bf $-\frac{1}{2}<l<0$} & \resizebox{0.6cm}{!}{\includegraphics{top49}} & \resizebox{1cm}{!}{\includegraphics{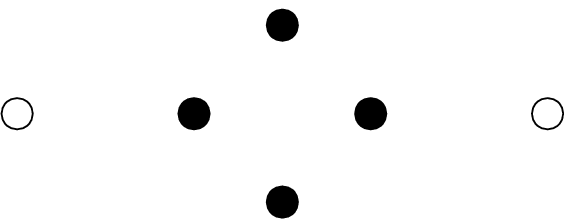}} & \resizebox{0.7cm}{!}{\includegraphics{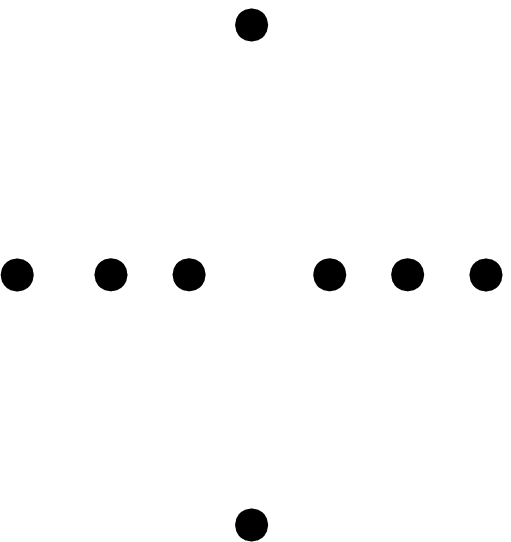}} & \resizebox{1cm}{!}{\includegraphics{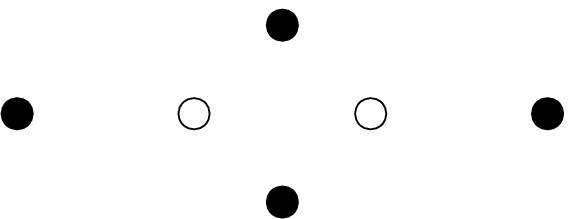}} & \resizebox{0.7cm}{!}{\includegraphics{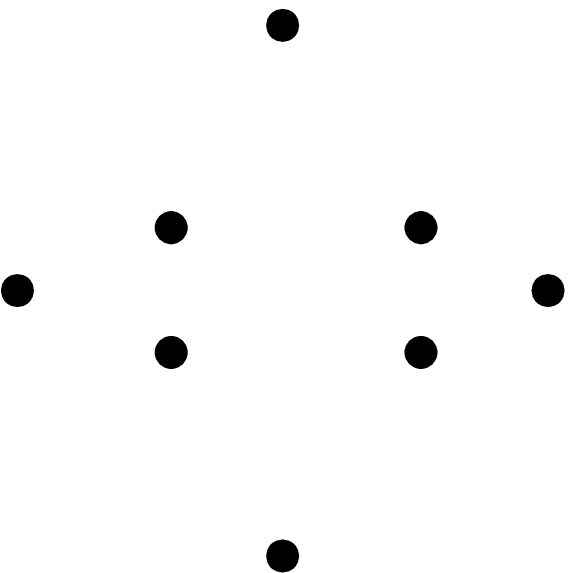}} & \resizebox{0.7cm}{!}{\includegraphics{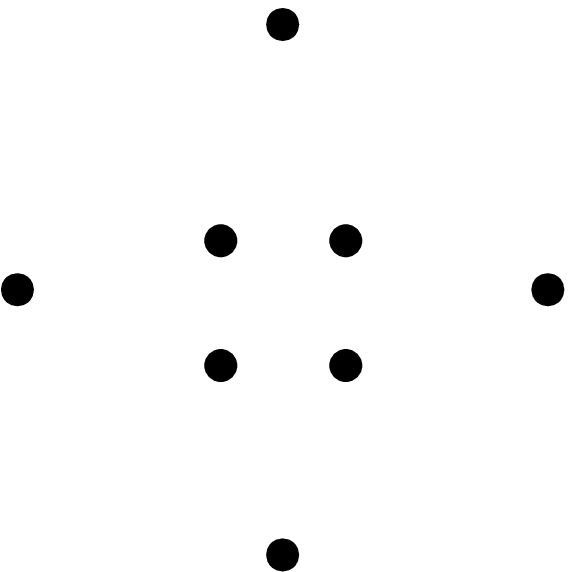}} & \resizebox{0.7cm}{!}{\includegraphics{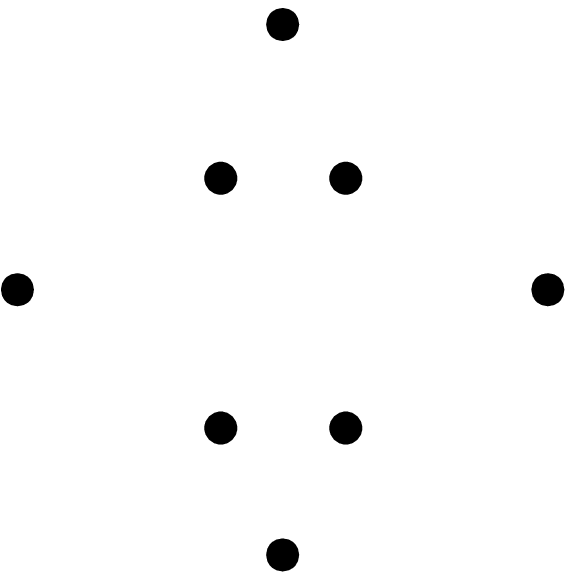}} &  \resizebox{0.3cm}{!}{\includegraphics{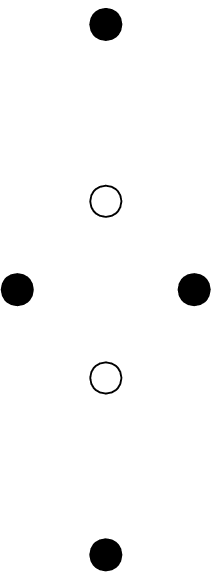}} & \resizebox{0.7cm}{!}{\includegraphics{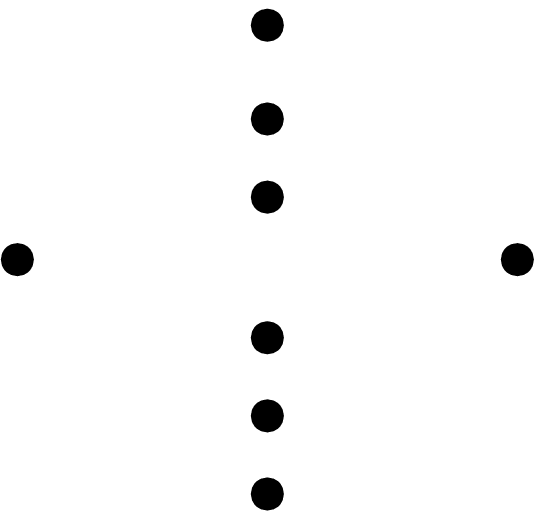}} & \resizebox{0.2cm}{!}{\includegraphics{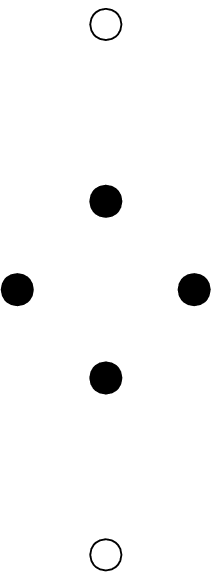}} & \resizebox{0.7cm}{!}{\includegraphics{top48}}  \\ \hline
\end{tabular}
\begin{tabular}{ccc}
 \\
\end{tabular}

\begin{tabular}{|c|c|c|c|c|c|c|c|c|c|}
\hline
\scriptsize Region & {\bf } & \scriptsize {\bf $E<-E^{\prime}$} & \scriptsize {\bf $E=-E^{\prime}$} & \scriptsize {\bf $-E^{\prime}<E<0$} & \scriptsize {\bf $E=0$} & \scriptsize {\bf $0<E<E^\prime$} & \scriptsize {\bf $E=E^{\prime}$} & \scriptsize {\bf $E>E^{\prime}$} \\ 
 \scriptsize (Figure \ref{domain1}) & & & & & & & & \\ \hline \hline
 \tiny {\bf Boundary of $E/F$} & \scriptsize {\bf $l=0$} & \resizebox{0.7cm}{!}{\includegraphics{top42}} & \resizebox{0.7cm}{!}{\includegraphics{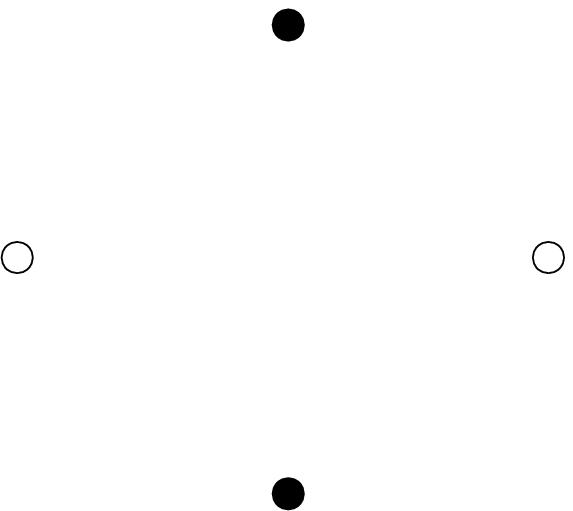}} & \resizebox{0.7cm}{!}{\includegraphics{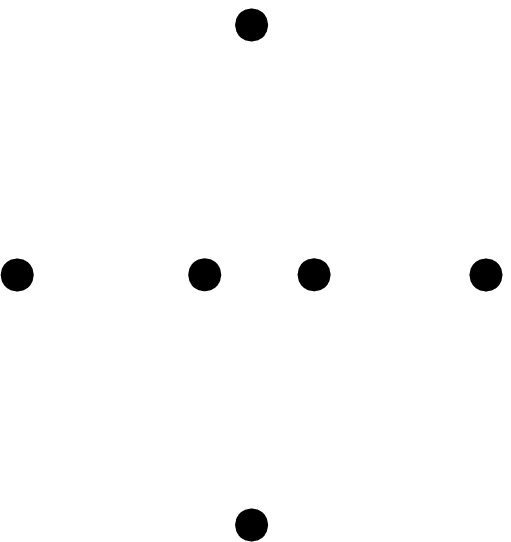}} & \resizebox{0.7cm}{!}{\includegraphics{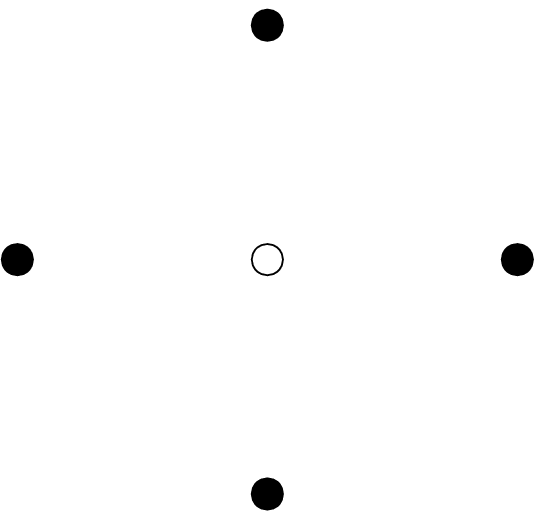}} & \resizebox{0.7cm}{!}{\includegraphics{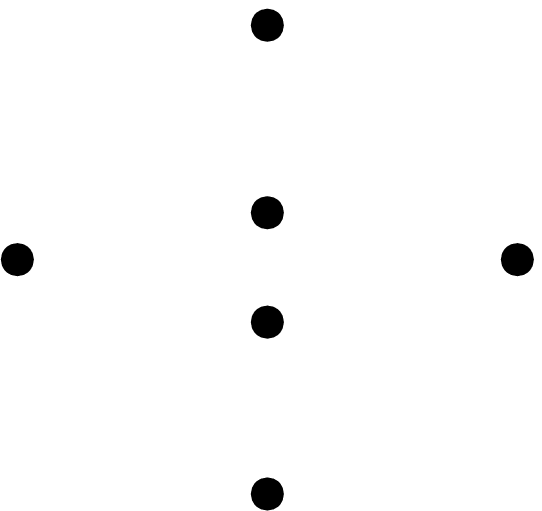}} & \resizebox{0.7cm}{!}{\includegraphics{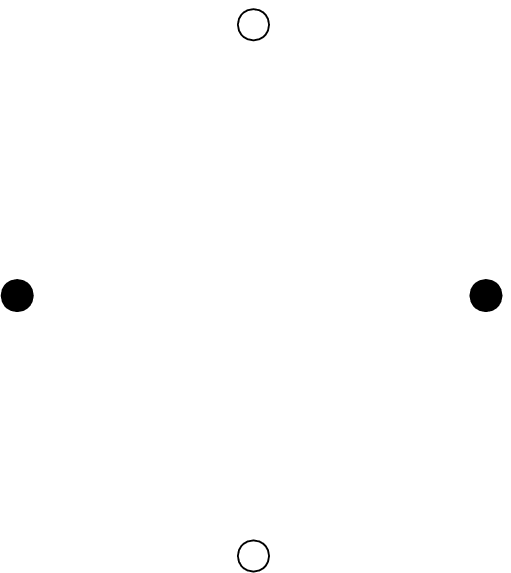}} & \resizebox{0.7cm}{!}{\includegraphics{top41}} \\ \hline
\end{tabular}
\end{center}
\end{table}
\footnotetext[3]{$-E^\prime=0$ for $l=l^\prime$}
\footnotetext[9]{$E^\prime=0$ for $l=l^\prime$}
\footnotetext[1]{Change in Stokes structure actually occurs at a value of $E$ slightly larger than $E^{\prime\prime}$.  Hence the quantisation condition associated with this arrangement does not come into effect exactly at $E=E^{\prime\prime}$}
\end{landscape}
\renewcommand{\thefootnote}{\arabic{footnote}}
\noindent To summarise the organisation of these different structures: for a given $\ga$, the sequence of turning point configurations obtained will depend on the value of $l$ relative to these key values.  This enables us to divide the $(\ga,l)$ plane up into sectors according to which sequence of turning point arrangements is valid there.  A sequence is obtained as the pattern of zeros obtained also depends on energy.   \\
The regions of the $(\ga,l)$ plane where different turning point configurations occur are shown in Figure \ref{domain1}.  This shows how the plane is divided in terms of these several key values of angular momentum, $l$.  The boundaries between these regions (shown as solid lines in Figure \ref{domain1}) are given by the lines $\ga=0$, $l=0$ and $l=\pm l^\prime$.  It is also apparent that the problem (\ref{eig}) is symmetric about $l=-\frac{1}{2}$. \\
Note that combinations of these sections approximate sections $A$ - $D$ of Figure \ref{domain0}, and the borders of the latter are included in Figure \ref{domain1} for comparison (dashed lines).  Later, we will compare our results with those found by Dorey {\it et al.} \cite{MR1857169,Dorey:2001hi}, and refer back to Figure \ref{domain0}.  (The reader may also note that although these regions do not match up perfectly with those found by Dorey {\it et al.}, all results are consistent;  using our method, reality and positivity are demonstrated for smaller regions than in \cite{MR1857169,Dorey:2001hi}, so we have a weaker condition overall - this point will be returned to in sections \ref{pos} and \ref{rea}). 
\begin{figure}[here]
\begin{center}
\resizebox{10.95cm}{!}{\includegraphics{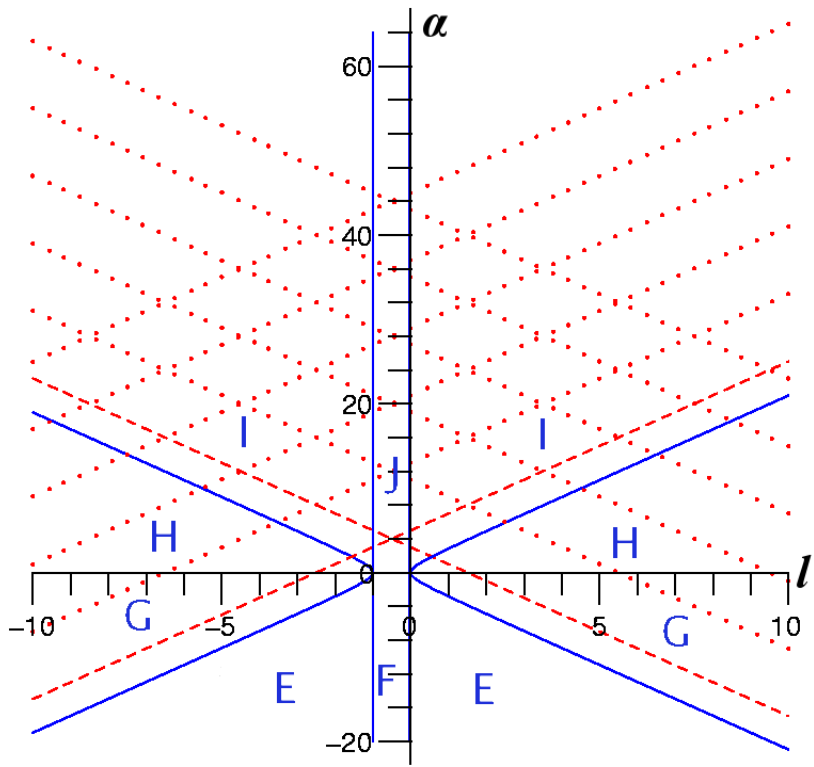}}
\caption{{\bf Regions of the $\ga$ - $l$ plane \label{domain1}}.\hspace{0.5cm}{\em The solid lines show how the plane is split up in terms of different Stokes structures occurring.  The dashed lines indicate the boundaries from  \cite{MR1857169,Dorey:2001hi}.}}
\end{center}
\end{figure}\\
Tables (\ref{plus}) and (\ref{minus}) are organised so that one first looks at the sign of $\ga$ to decide which table to use (Table \ref{plus} is for positive values of $\ga$), then which of the five rows to use depending on the value of $l$. Each row shows how the arrangement of zeros changes as $E$ is varied. \\
 Looking along one of these rows then, it can be seen that change occurs at $E=0$, or one of four different critical energy values (labelled $E^{\prime}$, $E^{\prime\prime}$, $E^{\prime\prime\prime}$ and $E_0$). $E^\prime$ and $E_0$ refer to energies at which turning points come together to form the double zeros described earlier, while $E^{\prime\prime}$ and $E^{\prime\prime\prime}$ are energy values at which turning points line up.  Expressions for each of these critical values (as a function of $\ga$ and $l$) have been obtained (though they are mostly unenlightening), and these energies can easily be calculated exactly for any particular values of the other parameters.
\subsubsection{Example of finding WKB quantisation condition}\label{examplesec}
\begin{figure}[here]
\begin{center}
\resizebox{12cm}{!}{\includegraphics{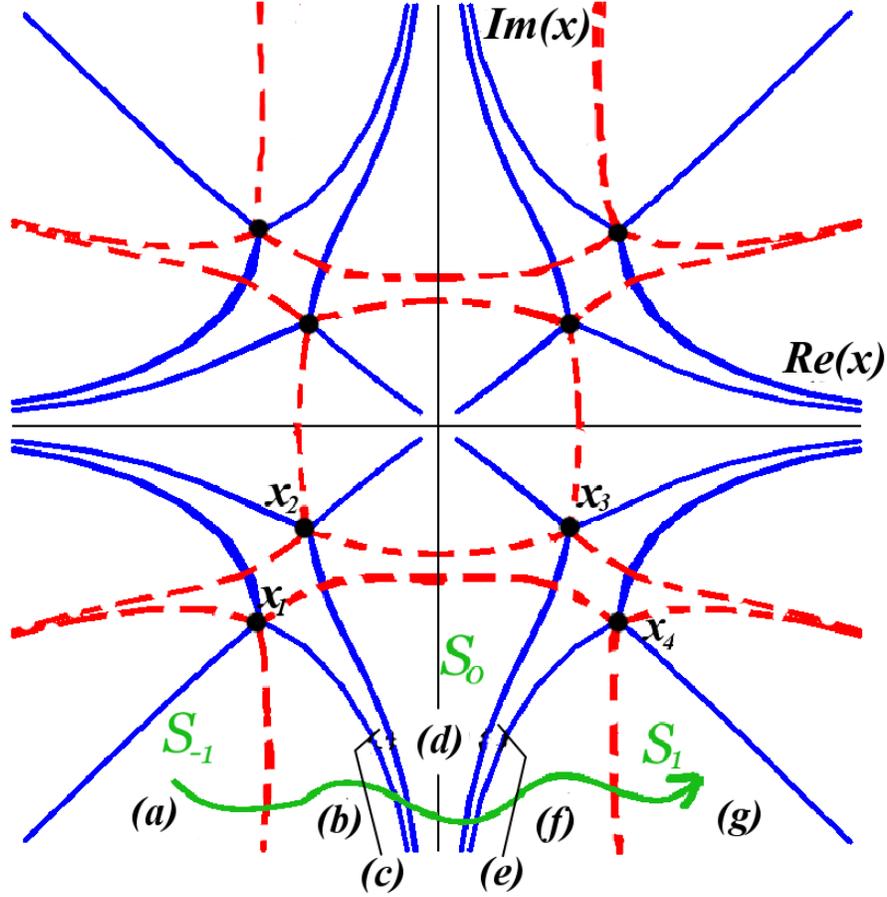}}
\caption{{\bf Stokes structure, $\ga=3$ , $l=0.5$ , $E=1$\label{example}} {\em Stokes lines are shown as solid lines and anti-Stokes lines are the broken lines.}}
\end{center}
\end{figure}

Figure \ref{example} shows the structure of Stokes and anti-Stokes lines in the case $\ga=3$, $l=0.5$ and $E=1$.  $l$ in this case is less than the critical value $l^\prime$ ($l^\prime \approx 1.081$ for $\ga=3$), so this refers to section $I$ in  Figure \ref{domain1} and Table \ref{plus}.  \\ \\
\\
In sector $\cS_{-1}$, the WKB wave function proportional to 
\bea
(P(x))^{-\frac{1}{4}}\mbox{e}^{i\omega(x_1,x)} \nn
\eea
is subdominant, where 
\bea
\omega(a,b)=  \int_a^b(P(t))^{\frac{1}{2}}dt =\int_a^b(E-V(t))^{\frac{1}{2}}dt \nn
\eea
and the $x_i$ are the turning points shown in Figure \ref{example}.  (In the following, dominance or subdominance of a solution is denoted by $(\bf d)$ or $(\bf s)$). We now continue this solution in an anticlockwise direction, and take account of new contributions appearing at Stokes lines and the dominance changes at anti-Stokes lines:\\ \\
\begin{math}
\displaystyle \mbox{(a)} \hspace{1cm}{ (P(x))^{-\frac{1}{4}}\mbox{e}^{i\omega(x_1,x)} \atop {\bf (s)}} \\ \\
\end{math}
Going from (a) to (b) now involves crossing an anti-Stokes line, so the subdominant solution now becomes dominant.  We will suppress the $ (P(x))^{-\frac{1}{4}}$ term in the remainder of the discussion for clarity. \\ \\
\begin{math}
\displaystyle \mbox{(b)} \hspace{1cm} {\mbox{e}^{i\omega(x_1,x)} \atop {\bf (d)}} \\ \\
\end{math}
To pass from region (b) to (c), the Stokes line coming from $x_1$ is crossed.  The dominant term remains unchanged, and the subdominant term gains an extra contribution equal to the dominant coefficient multiplied by the Stokes multiplier for this line ($i$).\\ \\
\begin{math}
\displaystyle \mbox{(c)} \hspace{1cm} {\mbox{e}^{i\omega(x_1,x)} \atop {\bf (d)}} {+ \atop }{i\mbox{e}^{-i\omega(x_1,x)}\atop {\bf (s)}} \\ \\
\end{math}
This solution is re-written in terms of the WKB solution from turning point $x_2$.\\ \\
\begin{math}
\displaystyle \hspace*{0.95cm}={\mbox{e}^{i\omega(x_1,x_2)}\mbox{e}^{i\omega(x_2,x)} \atop {\bf (d)}} {+ \atop }{i\mbox{e}^{-i\omega(x_1,x_2)}\mbox{e}^{-i\omega(x_2,x)} \atop {\bf (s)}}\\ \\
\end{math}
Next a Stokes line from $x_2$ is crossed, and so the subdominant term again picks up a contribution proportional to the dominant term. \\ \\
\begin{math}
\displaystyle \mbox{(d)} \hspace{1cm} {\mbox{e}^{i\omega(x_1,x_2)}\mbox{e}^{i\omega(x_2,x)} \atop {\bf (d)}} { + \atop } {i\left[\mbox{e}^{-i\omega(x_1,x_2)}+\mbox{e}^{i\omega(x_1,x_2)}\right]\mbox{e}^{-i\omega(x_2,x)} \atop {\bf (s)}} \\ \\ \end{math}
This is again written in terms of the WKB solution from $x_3$, in preparation for crossing the Stokes line from that turning point.\\ \\
\begin{math}
\hspace*{0.95cm}={\mbox{e}^{i\omega(x_1,x_2)}\mbox{e}^{i\omega(x_2,x_3)}\mbox{e}^{i\omega(x_3,x)} \atop {\bf (d)}}{+ \atop }{i\left[\mbox{e}^{-i\omega(x_1,x_2)}+\mbox{e}^{i\omega(x_1,x_2)}\right]\mbox{e}^{-i\omega(x_2,x_3)}\mbox{e}^{-i\omega(x_3,x)} \atop {\bf (s)}}\\ \\
\end{math}
This procedure repeats, as a further Stoke line is crossed.\\ \\
\begin{math}
\mbox{(e)} \hspace{1cm}{\mbox{e}^{i\omega(x_1,x_2)}\mbox{e}^{i\omega(x_2,x_3)}\mbox{e}^{i\omega(x_3,x)}  \atop {\bf (d)}} {+ \atop } {i\left\{\left[\mbox{e}^{-i\omega(x_1,x_2)}+\mbox{e}^{i\omega(x_1,x_2)}\right]\mbox{e}^{-i\omega(x_2,x_3)}+\mbox{e}^{i\omega(x_1,x_2)}\mbox{e}^{i\omega(x_2,x_3)}\right\}\mbox{e}^{-i\omega(x_3,x)} \atop {\bf (s)}} \\ \\
\hspace*{0.95cm}={\mbox{e}^{i\omega(x_1,x_2)}\mbox{e}^{i\omega(x_2,x_3)}\mbox{e}^{i\omega(x_3,x_4)}\mbox{e}^{i\omega(x_4,x)} \atop {\bf (d)}} \\ \\
\hspace*{1.95cm}{+ \atop }{i\left\{\left[\mbox{e}^{-i\omega(x_1,x_2)}+\mbox{e}^{i\omega(x_1,x_2)}\right]\mbox{e}^{-i\omega(x_2,x_3)}+\mbox{e}^{i\omega(x_1,x_2)}\mbox{e}^{i\omega(x_2,x_3)}\right\}\mbox{e}^{-i\omega(x_3,x_4)}\mbox{e}^{-i\omega(x_4,x)} \atop {\bf (s)}} \\ \\
\mbox{(f)} \hspace{1cm}{\mbox{e}^{i\omega(x_1,x_2)}\mbox{e}^{i\omega(x_2,x_3)}\mbox{e}^{i\omega(x_3,x_4)}\mbox{e}^{i\omega(x_4,x)} \atop {\bf (d)}} \\ \\
\hspace*{1.95cm} {+ i\left[\left\{\left[\mbox{e}^{-i\omega(x_1,x_2)}+\mbox{e}^{i\omega(x_1,x_2)}\right]\mbox{e}^{-i\omega(x_2,x_3)}+\mbox{e}^{i\omega(x_1,x_2)}\mbox{e}^{i\omega(x_2,x_3)}\right\}\mbox{e}^{-i\omega(x_3,x_4)} \right. \atop } \\ \\
\hspace*{2.95cm}{ + \atop}{\left. \mbox{e}^{i\omega(x_1,x_2)}\mbox{e}^{i\omega(x_2,x_3)}\mbox{e}^{i\omega(x_3,x_4)}\right]\mbox{e}^{-i\omega(x_4,x)} \atop {\bf (s)}} \\ \\
\end{math}
Finally, in going from (f) to (g), an anti-Stokes line is crossed again, and dominance is again exchanged.\\ \\
\begin{math}
\displaystyle \mbox{(g)} \hspace{1cm}{\mbox{e}^{i\omega(x_1,x_2)}\mbox{e}^{i\omega(x_2,x_3)}\mbox{e}^{i\omega(x_3,x_4)}\mbox{e}^{i\omega(x_4,x)} \atop {\bf (s)}} \\ \\
\hspace*{1.95cm}{+i\left[\left\{\left[\mbox{e}^{-i\omega(x_1,x_2)}+\mbox{e}^{i\omega(x_1,x_2)}\right]\mbox{e}^{-i\omega(x_2,x_3)}+\mbox{e}^{i\omega(x_1,x_2)}\mbox{e}^{i\omega(x_2,x_3)}\right\}\mbox{e}^{-i\omega(x_3,x_4)} \right. \atop } \\ \\
\hspace*{2.95cm}{+ \atop }{\left. \mbox{e}^{i\omega(x_1,x_2)}\mbox{e}^{i\omega(x_2,x_3)}\mbox{e}^{i\omega(x_3,x_4)}\right]\mbox{e}^{-i\omega(x_4,x)} \atop {\bf (d)}} \\ \\
\end{math}
So for the solution to be purely subdominant at (g), (Sector $\cS_{1}$), in order to satisfy the boundary condition, we require
\bea
\displaystyle \left\{\left[\mbox{e}^{-i\omega(x_1,x_2)}+\mbox{e}^{i\omega(x_1,x_2)}\right]\mbox{e}^{-i\omega(x_2,x_3)}+\mbox{e}^{i\omega(x_1,x_2)}\mbox{e}^{i\omega(x_2,x_3)}\right\}\mbox{e}^{-i\omega(x_3,x_4)} + \mbox{e}^{i\omega(x_1,x_2)}\mbox{e}^{i\omega(x_2,x_3)}\mbox{e}^{i\omega(x_3,x_4)}=0 \nn \\ \nn
\eea
Writing $\omega(x_1,x_2)=U+iV=(\omega(x_3,x_4))^*$ and $\omega(x_2,x_3)=W$, with $U,V$ and $W$ real, ($\omega(x_2,x_3)$ can be seen to be purely real as $x_2$ and $x_3$ are joined by an anti-Stokes line), following the example of \cite{BenderBerry:2001wk}, this simplifies to:
\bea
E \mbox{ eigenvalue} \iff 2\cos(2U+W) + 2\mbox{e}^{-2V}\cos(W) &=& 0 \nn
\eea
As explained earlier, the form of the quantisation condition changes, depending on the parameters $\ga$ and $l$, as well as changing as energy is increased moving along a sequence.  We now define (piecewise), an overall quantisation condition function, which we will denote $Q(E,\ga,l)$.  {\it e.g.}  as we have just seen $Q(E,\ga,l)= 2\cos(2U+W) + 2\mbox{e}^{-2V}\cos(W)$ for $0<E<E^\prime$ , $\ga>0$ , $0<l<l^\prime$.  Similarly $Q(E,\ga,l)$ is defined to be the quantisation condition obtained for each of the other different sets of parameter values. Finding eigenvalues is then a matter of finding the zeros of this function.
\subsection{Positivity of Spectra from WKB Quantisation Conditions}\label{pos}
The sections $E$ and $F$ in Figure \ref{domain1} form an area that is completely contained in the sector where positivity was proved in \cite{MR1857169} (bounded by the red lines in Figure \ref{domain1}, or section D in Figure \ref{domain0}).\\
WKB analysis of $E$ and $F$ shows that no quantisation condition can exist for negative energies.  In fact, a viable quantisation condition is only found for $E>E^\prime$.  For energy values lower than this, demanding that the dominant wave function vanishes in the Stokes sector  $\cS_{1}$ also means that the subdominant wave function will vanish; there is no way to have a WKB solution that is purely subdominant in sectors $\cS_{-1}$ and  $\cS_{1}$, and hence there are no eigenvalues.
\subsection{Conjecture on Reality of Spectra from WKB Quantisation Conditions}\label{rea}
\begin{figure}[here]
\begin{center}
\subfigure[]
{\label{qua}
\resizebox{5.1cm}{!}{\includegraphics{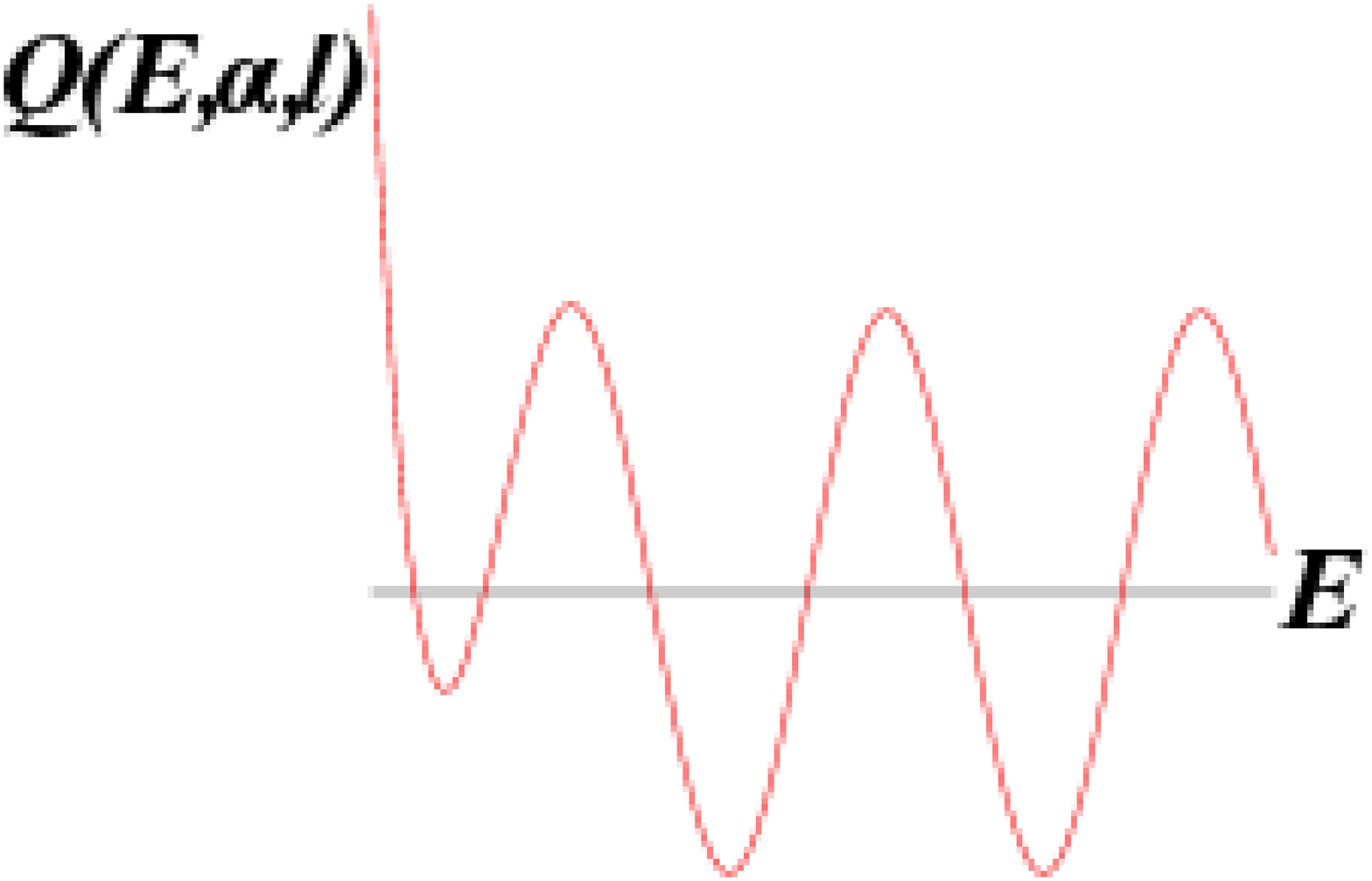}}} \hspace{0.5cm}
\subfigure[]
{\label{qub}
\resizebox{5.1cm}{!}{\includegraphics{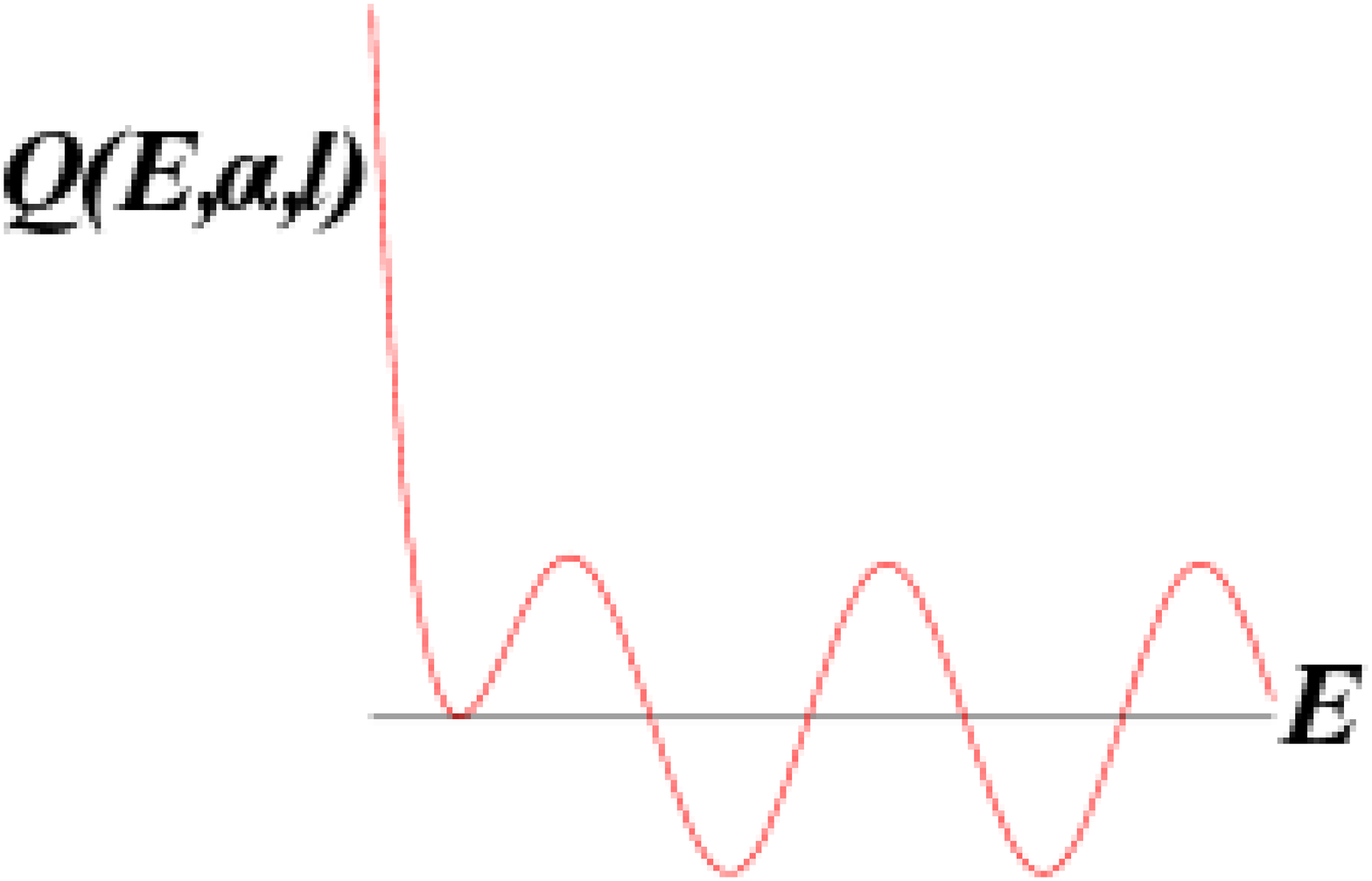}}} \hspace{0.5cm}
\subfigure[]
{\label{quc}
\resizebox{5.1cm}{!}{\includegraphics{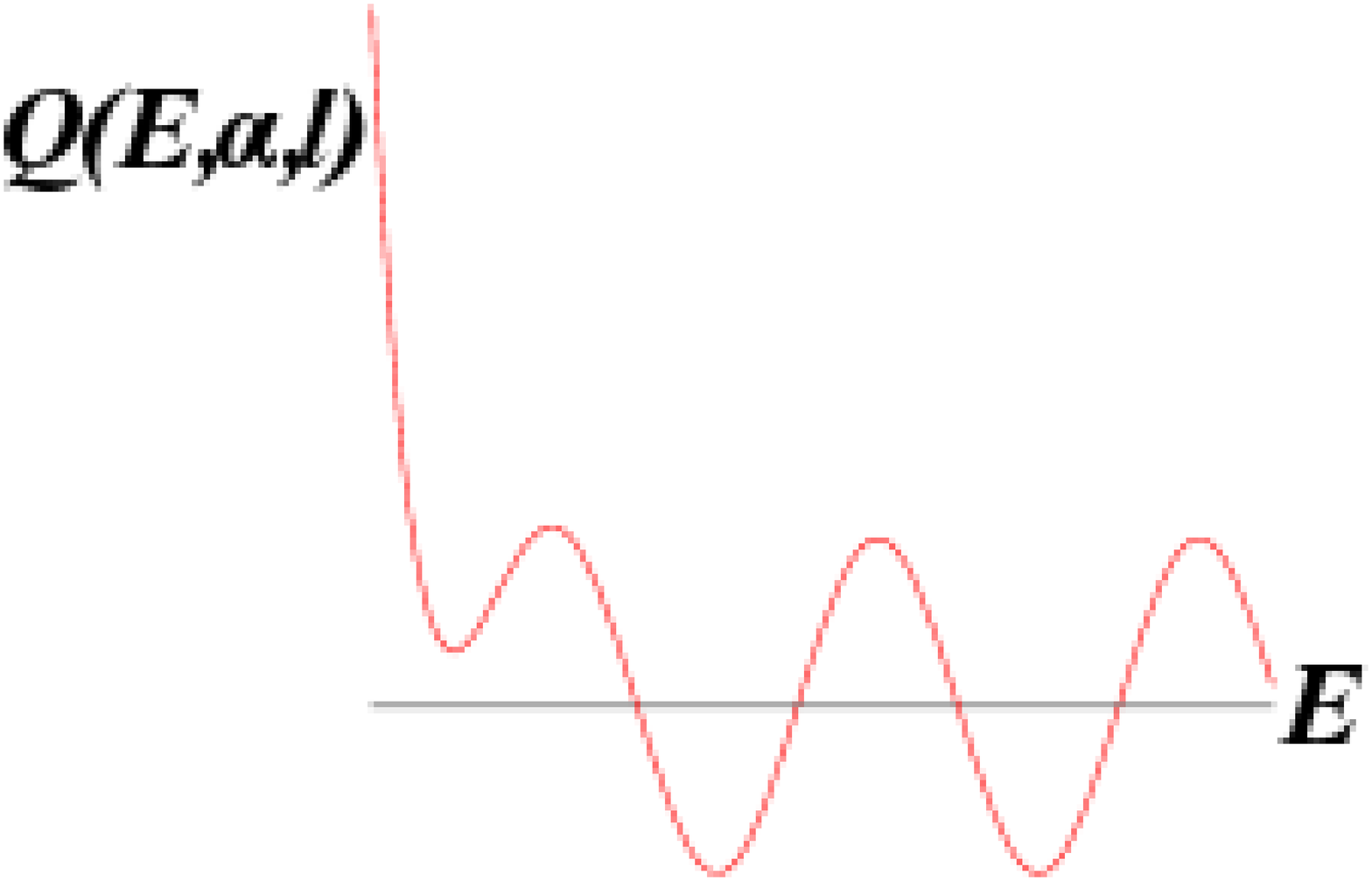}}}
\caption{{\bf Quantisation condition function plotted against energy} {\em \ref{qua} corresponds to 2 real eigenvalues, \ref{qub} to real, degenerate eigenvalues and \ref{quc} to complex conjugate eigenvalues.  (Referring to two lowest energy eigenvalues shown).}\label{qmin}}
\end{center}
\end{figure}

\begin{figure}[here]
\begin{center}
\subfigure[]
{\label{qumin}
\resizebox{6.8cm}{!}{\includegraphics{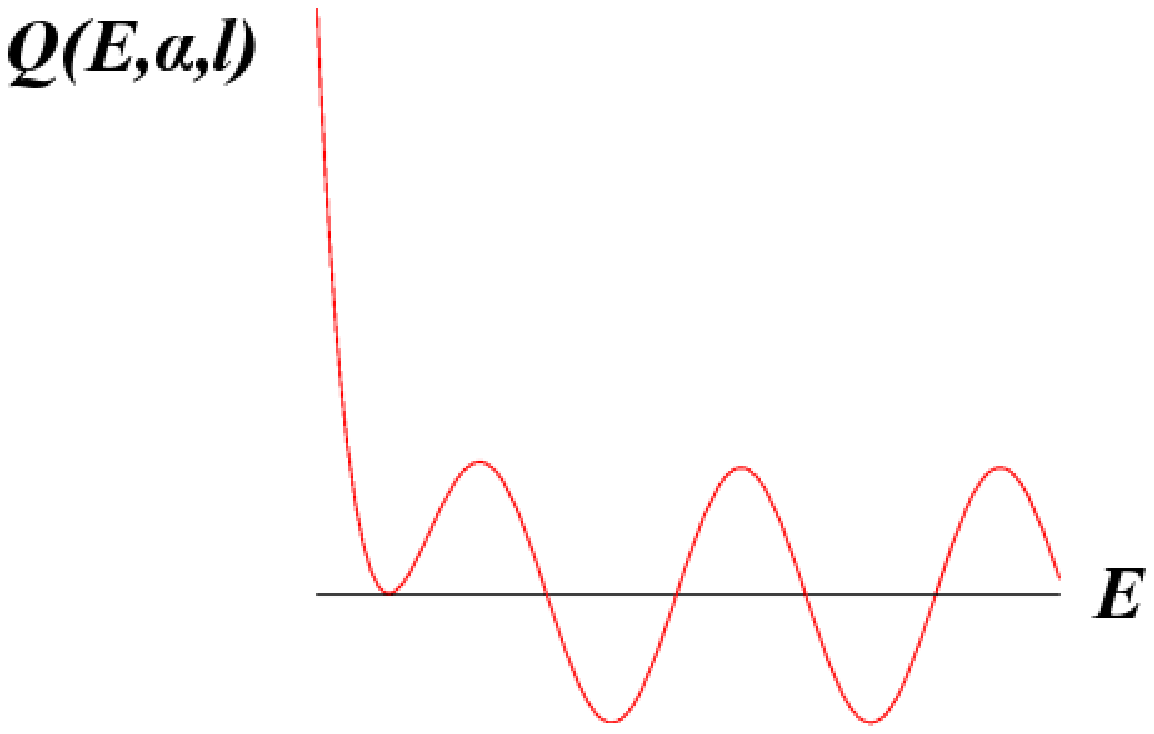}}} \hspace{0.5cm}
\subfigure[]
{\label{qumax}
\resizebox{6.8cm}{!}{\includegraphics{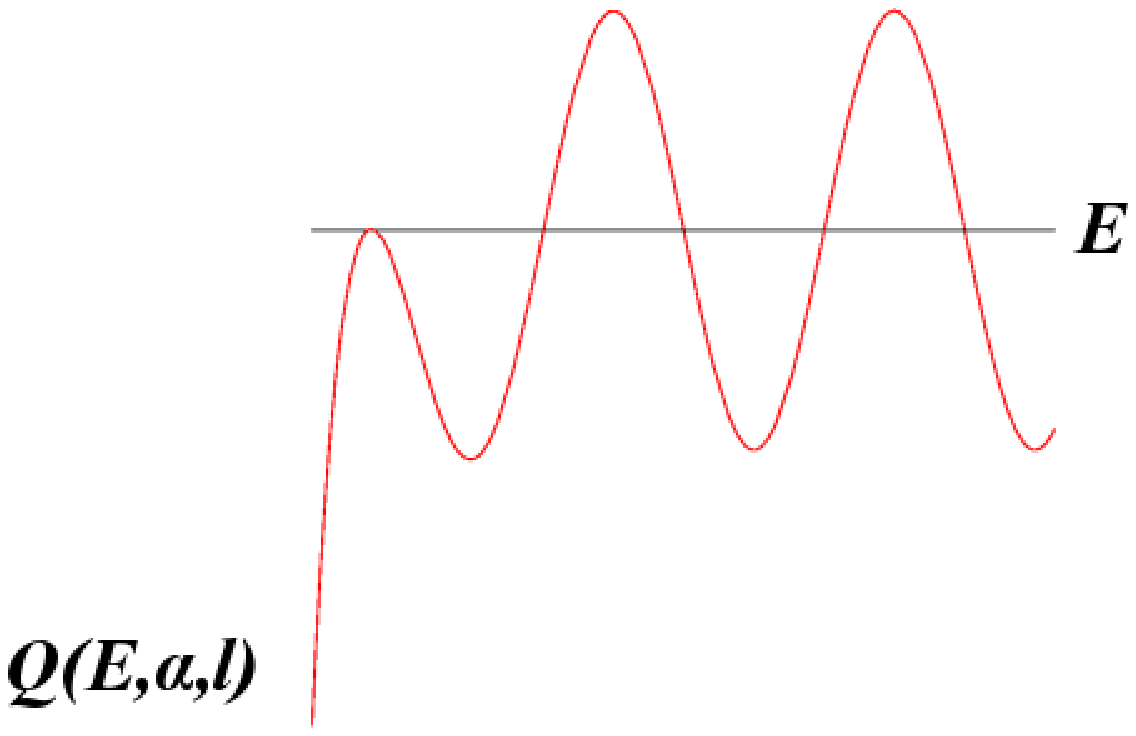}}}
\caption{{\bf Quantisation condition function, }{\em illustrating degeneracies arising from mimimum being pulled up, and maximum being pulled down.}\label{qminmax}}
\end{center}
\end{figure}

In \cite{BenderBerry:2001wk}, the appearance of complex eigenvalues was explained in terms of a WKB quantisation condition.  In the paper, Bender {\it et al.} obtained a WKB quantisation condition for the $\pt$ symmetric potential
\bea
V(x)=x^4+iAx. \nn
\eea
The condition found was:
\bea
E \mbox{ eigenvalue} \iff \cos(2U)+\frac{1}{2}\mbox{e}^{-2V}=0, \nn
\eea
\bea
\mbox{where } U=\Re e \left(\int_{x_1}^{x_3}(E-V(t))^{\frac{1}{2}}dt\right), \hspace{0.5cm} V= \Im m\left(\int_{x_1}^{x_3}(E-V(t))^{\frac{1}{2}}dt\right) \nn
\eea
This condition, consisting of the sum of an exponential term and an oscillatory term can lead to real eigenvalues, provided that the magnitude of the exponential term does not exceed $1$.  If parameters are varied, leading to an increase in magnitude of the exponential term beyond $1$, then no real solutions are possible.  In effect, real eigenvalues change in value as the parameter $A$ is varied continuously, then `pair off' and disappear as the exponential term becomes larger.  This can be seen explicitly by plotting the value of the quantisation condition function against energy for a particular value of parameter $A$, then looking at how this plot changes with different values of $A$.  As the value of the exponential term increases with changing A, local minima in the plot can be seen to be pulled up through zero, leading to degenerate eigenvalues and complex conjugate eigenvalues.  (see Figure \ref{qmin}).
\\
The WKB quantisation conditions found in regions $I$ and $J$ of Figure \ref{domain1}, (which are slightly larger than the region $A$ (Figure \ref{domain0}) discussed in \cite{MR1857169,Dorey:2001hi}), were found to be similar in form to that obtained in \cite{BenderBerry:2001wk} ({\it i.e.} containing an exponential and oscillatory term).\\
With these conditions the effect described above can occur, where pairs of real eigenvalues move together and become degenerate as the parameters change, due to the effect of the exponential term(s) in the quantisation condition. \\
An example of this is seen by looking at the quantisation conditions obtained for Region $I$ (Table \ref{plus}).  In section \ref{examplesec} we showed that the quantisation condition for region $I$ ($\ga>0,0<l<l^\prime$) with $0<E<E^\prime$ is
\bea
E \mbox{ eigenvalue} \iff 2\cos(2U+W) + 2\mbox{e}^{-2V}\cos(W) &=& 0 \nn
\eea
Plotting this quantisation condition function against energy, for various values of $\ga$ and $l$, it can be seen that varying the parameters can again lead to the exponential term  increasing in magnitude so that it begins to dominate the oscillatory term, and pulls the function through zero.  One important difference to \cite{BenderBerry:2001wk} is that the exponential term in the condition is multiplied by a cosine term.  This makes it harder to analytically pin down the boundary in the parameter space where degenerate eigenvalues may occur, than  in \cite{BenderBerry:2001wk}.  (Hence we can only restrict the domain of unreality to the regions where particular quantisation conditions are valid, and don't, as yet, make a more detailed study of these regions - {\it i.e.} we only say that complex conjugate eigenvalues exist somewhere in this region, and do not analytically calculate the particular area inside this region in which real eigenvalues are not possible (c.f. \cite{BenderBerry:2001wk})). \\
This cosine term leads to another difference in the quantisation condition function plots; because this term varies between $\pm 1$, the exponential term in the condition will be positive for some values of $\ga,l,E$ and negative for others.  degenerate eigenvalues can now come about through a local minimum being pulled up through zero, or a local maximum being pulled down.  See Figure \ref{qminmax}.\\\\
\\\\
In all other regions, the quantisation conditions do not contain exponential terms.   The most common quantisation condition found outside of I and J is of the form $\cos (U) =0$, apparently leading to an infinite number of real eigenvalues.  We conjecture that the only complex eigenvalues found for this problem are the ones that come about in the manner described earlier, whereby the exponential term allows for `pulling through' of the quantisation condition function.  If it is true, as seems to be the case numerically, that this is the only means of complex eigenvalues coming into existence, then any quantisation condition lacking an exponential term (or at least a sum involving two similar sized terms) will not lead to complex eigenvalues.   Hence our conjecture leads to the conclusion that, in Figure \ref{domain1}, the spectrum will be entirely real below the blue lines in the upper part of the plane.  This is a weaker condition than that found in \cite{MR1857169}, where reality was established for all parameter values below the red lines in this upper part of the plane. \\
(Note that there is one exception to the statement that the quantisation condition does not contain exponential terms outside of I and J; that being the  case $l>0,E>E^{\prime\prime}$, {\it i.e.} high energy values in regions $E,G,H$ and $I$;  The quantisation condition found here is of the same form, $2\cos(2U+W) + 2\mbox{e}^{-2V}\cos(W) = 0$, but this condition only comes into effect for $E>E^{\prime\prime}$ whence it is seen numerically that $V$ is relatively large and positive, so the exponential term is approximately zero and the pairing off of eigenvalues cannot occur, and so again only real eigenvalues are possible).
 \subsection{WKB explanation for appearance of degeneracies and cusps}\label{deg}
\begin{figure}[here!]
\begin{center}
\resizebox{13.8cm}{!}{\includegraphics{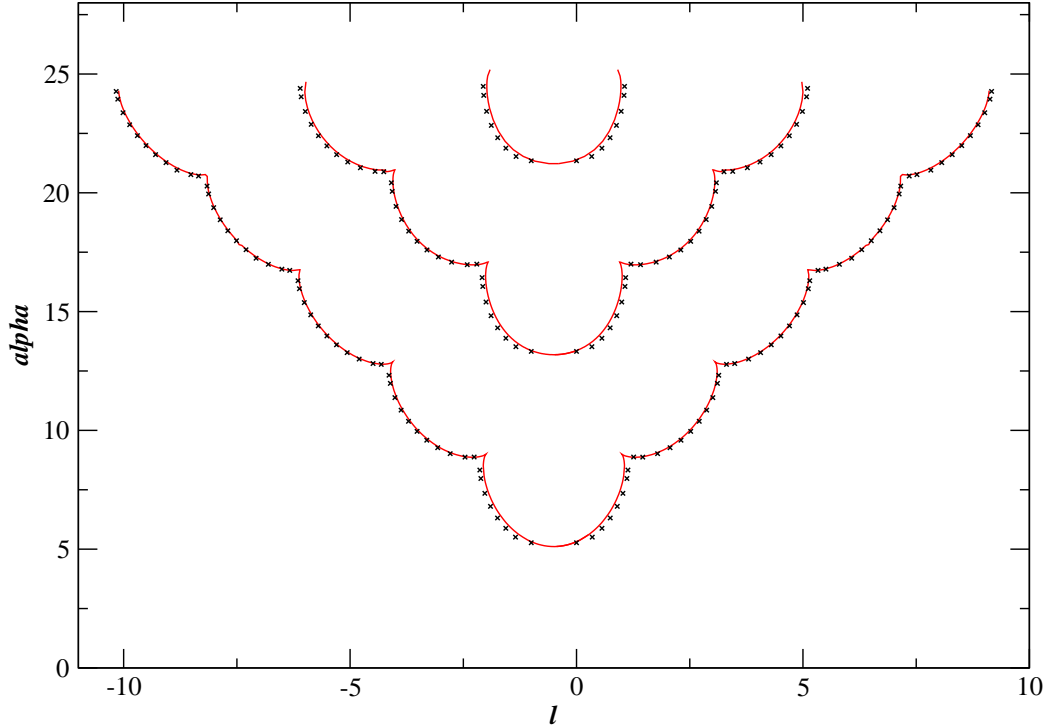}}
\caption{{\bf Degenerate eigenvalues} {\em The positions of degenerate eigenvalues found using WKB quantisation conditions (crosses) and those found with the methods of \cite{MR1857169,Dorey:2001hi} (lines).}\label{cuspa}}
\end{center}
\end{figure}

\begin{figure}[here!]
\begin{center}
\resizebox{8cm}{!}{\includegraphics{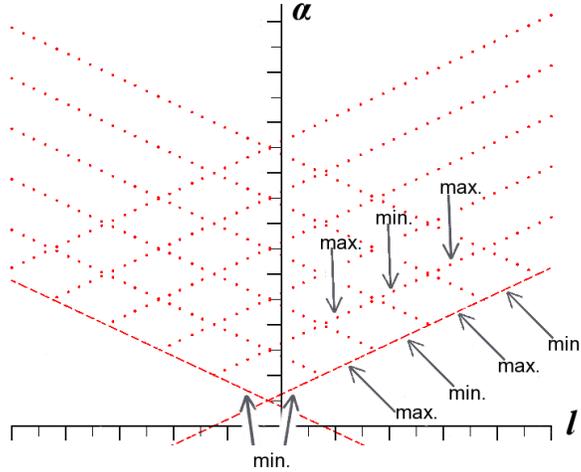}}
\caption{{\bf Degeneracies being formed from minima or maxima of $Q(E,\ga,l)$} {\em On the edge of one diamond shaped region, degeneracies occur as either a minimum or maximum.  In the neighbouring diamond this is swapped.}\label{domainf}}
\end{center}
\end{figure}

\begin{figure}[here!]
\begin{center}
\resizebox{7.2cm}{!}{\includegraphics{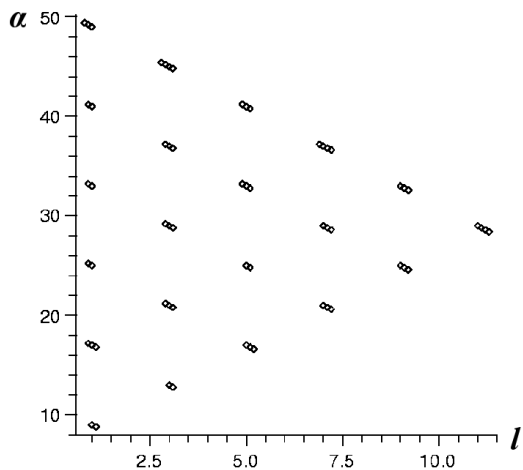}}
\caption{{\bf Locations of cusps. \label{cusps}}}
\end{center}
\end{figure}

The pattern of cusps in the degeneracy curves described in \cite{Dorey:2001hi} is manifest from the WKB approach via the quantisation conditions.  The position of degenerate eigenvalues can be found numerically by searching for the parameter values where a local maximum (minimum) in the quantisation condition passes through $0$.  this is done by combining a simple maximum (minimum) finding algorithm, and a bisection algorithm to find when the maximum (minimum) value of the quantisation condition function is $0$. In this way, a plot of the locations of degenerate eigenvalues can be found. (Figure \ref{cuspa}).\\ 
Figure \ref{cuspa} shows the pattern of curves, meeting at cusps, as seen in Figure \ref{scan}, first shown by Dorey {\it et al.}  \cite{MR1857169}.  Note that this figure is different to Figure \ref{scan}, as it shows the cusp pattern repeating itself as further eigenvalues join up and then split into complex conjugate pairs.  The results from the complex WKB quantisation conditions are shown as the crosses, overlaying the results from the method used in \cite{MR1857169,Dorey:2001hi}.  It can be seen that there is extremely good agreement between the degeneracy curves found via these two methods.  We are extremely grateful to  Patrick Dorey, Clare Dunning, Anna Lishman and Roberto Tateo for sharing their results prior to publication \cite{Dorey:unpub, Dorey:unpub2}.
\\
To explain how the quantisation condition method accounts for the formation of the cusps in the diagram, consider the diamond-shaped area of the $(\ga,l)$ plane bordered by $0<\ga_{-}<1$ and $0<\ga_{+}<1$, In this region, for a fixed value of $\ga_{+}$, and $\ga_{-}$ being increased from $0$ to $1$, degeneracies are seen as a local minimum of the quantisation condition function being pulled up through zero.  In the adjoining diamond-shaped area, $0<\ga_{-}<1$ and $1<\ga_{+}<2$, degeneracies occur as a local maximum is pulled down through zero.  In $0<\ga_{-}<1$ and $2<\ga_{+}<3$, it is again a minimum, and so on (Figure \ref{domainf}).
\\
On the boundaries of these regions, ie $\ga_{\pm} \in \Z^{+}$, the degeneracies occur as a limit of a local minimum and local maximum of the quantisation condition function both meeting at $Q(E,\ga,l)=0$, ie a point of inflection of the quantisation condition function.  These correspond to the cusps in \cite{Dorey:2001hi}.\\
\begin{figure}[here]
\begin{center}
\subfigure[]
{\label{qinf1}
\resizebox{3.6cm}{!}{\includegraphics{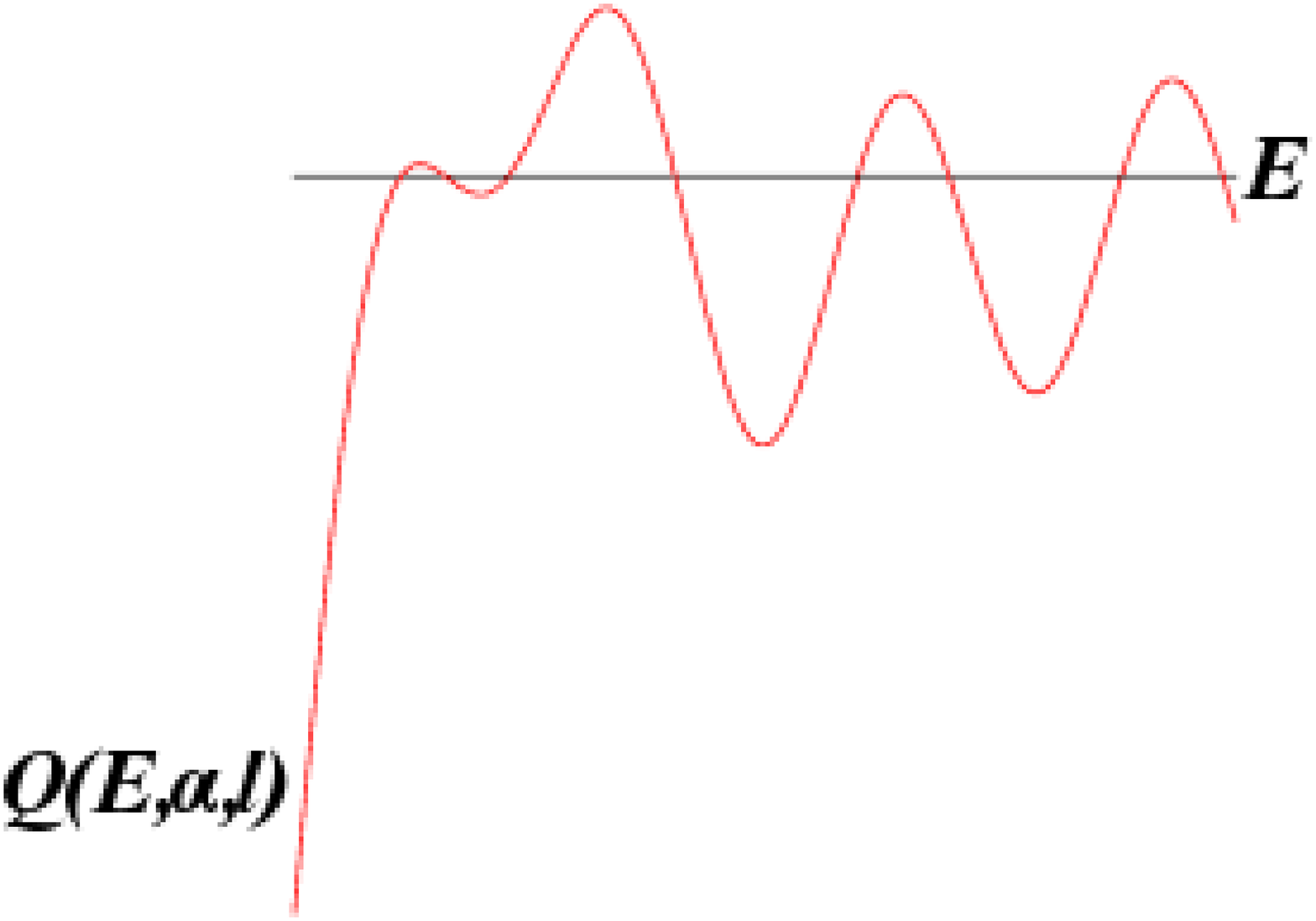}}} \hspace{0.5cm}
\subfigure[]
{\label{qinf2}
\resizebox{3.6cm}{!}{\includegraphics{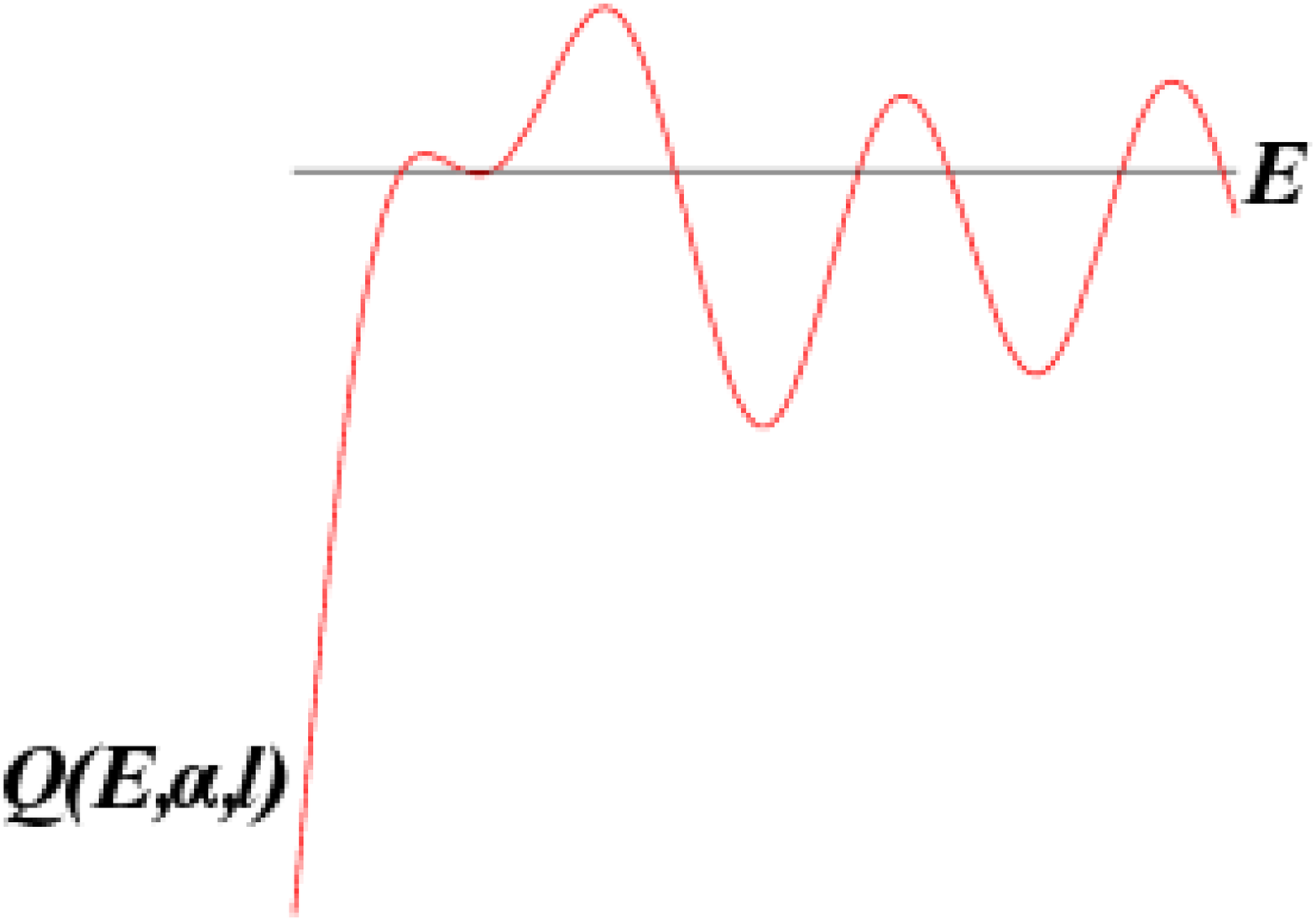}}} \hspace{0.5cm}
\subfigure[]
{\label{qinf3}
\resizebox{3.6cm}{!}{\includegraphics{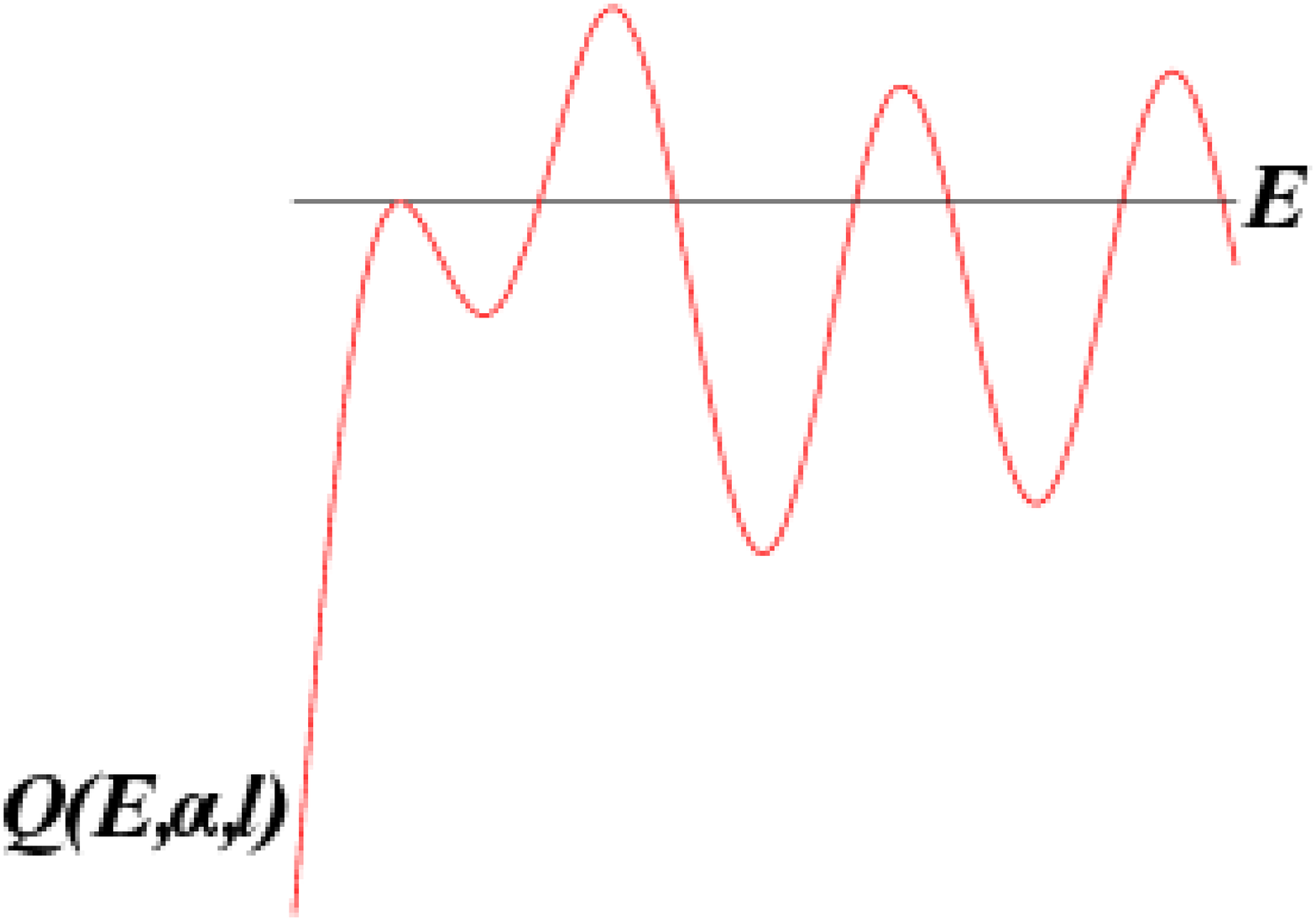}}} \hspace{0.5cm}
\subfigure[]
{\label{qinf4}
\resizebox{3.6cm}{!}{\includegraphics{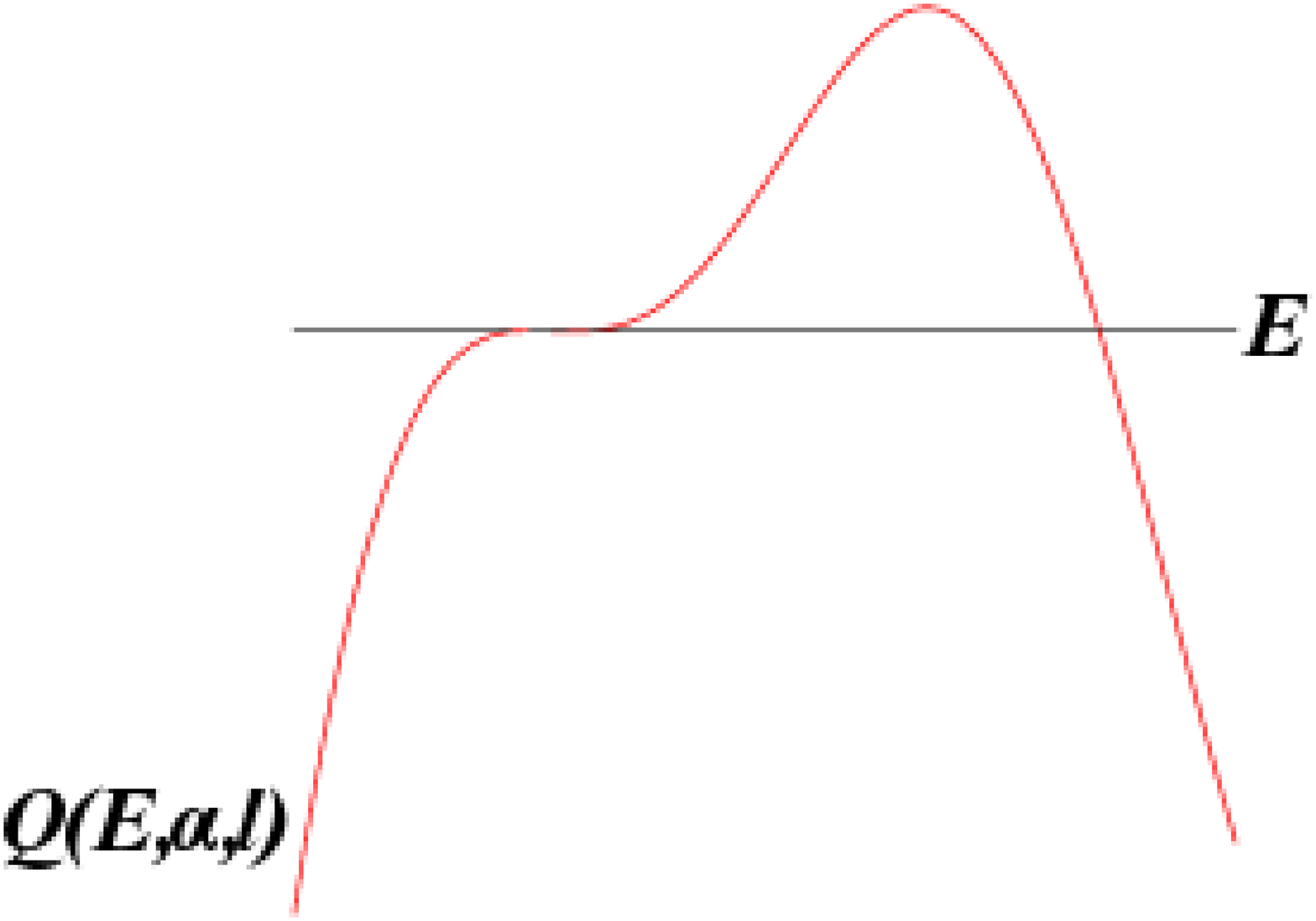}}}
\caption{{\bf Quantisation condition function, parameter values close to inflection point.}\label{qinf}}
\end{center}
\end{figure}\\
An example is shown in Figure \ref{qinf}.  \ref{qinf1} shows the behaviour of $Q$ with parameter values approaching those for which a point of inflection occurs.  It can be seen that both a local maximum and local minimum are close to $Q=0$.  Figure \ref{qinf2} corresponds to a degeneracy for $\ga_{+}$ slightly less than $2$, ($\ga_{-}\approx 1$).  Here the minimum has just been pulled up to $Q=0$.  \ref{qinf3} corresponds to a degeneracy for $\ga_{+}$ just greater than $2$, ($\ga_{-}\approx 1$ still).  Here the maximum has just been pulled down.  \ref{qinf4} shows $Q$ for the parameter values $\ga_{+}\approx2,\ga_{-}\approx 1$, leading to an inflection point.\\
Numerically, by scanning the $(\ga,l)$ plane for points of inflection, a plot of the positions of these cusps, as predicted by WKB was obtained (Figure \ref{cusps}). 
\\
These cusp positions are in approximate agreement with those found by Dorey {\it et al.} in \cite{MR1857169,Dorey:2001hi}.
\begin{section}{Conclusions and Further Directions}\label{conc}
We have shown that the approach of complex WKB quantisation conditions can be used to explain aspects of the spectrum of (\ref{eig}).  The problem appears to be more complicated due to the changing of the Stokes structures as parameters are varied, but in fact it is this changing of configurations that provides an explanation for the different behaviour in different regions of the $(\ga,l)$ plane. \\
The $M=3$ case is a useful test case to investigate as some energy levels can be found exactly, due to the quasi-exactly solvability of the related problem discussed in \cite{Dorey:2001hi}.  The quantisation condition method has been shown to give excellent agreement in calculating the positions of the degenerate eigenvalues, (possibly surprisingly good, as first order complex WKB is supposed to be a more accurate approximation for higher energy levels, and the pairing off occurs between low lying energy levels).  There should be no reason why the method would not be useful in investigating the spectrum for different values of $M$.  Indeed, considering smaller values of $M$ would be expected to be a simpler problem, as less zeros of the potential should lead to less complexity in the pattern of Stokes structures obtained.   It may be possible to see how the quantisation condition approach changes for general $M$, and gain insight into the way the domain of unreality changes as $M$ is varied.
\end{section}
\begin{subsubsection}*{Acknowledgements}
The author is very grateful to Robert Weston for many helpful ideas and much advice.  He would also like to to thank  Patrick Dorey, Clare Dunning, Chris Howls, Adri Olde Daalhuis, Michael Graham, Katie Russell and Des Johnston for useful discussions.  This work was funded by a James Watt scholarship.  Partial support was provided by the European TMR network EUCLID (contract number HPRN-CT-2002-00325).
\end{subsubsection}
\newpage

\begin{thebibliography}{10}

\bibitem{MR1857169}
Patrick Dorey, Clare Dunning, and Roberto Tateo.
\newblock Spectral equivalences, {B}ethe ansatz equations, and reality
  properties in {$\mathcal{PT}$}-symmetric quantum mechanics.
\newblock {\em J. Phys. A}, 34(28):5679--5704, 2001, [hep-th/0103051].

\bibitem{Dorey:2001hi}
Patrick Dorey, Clare Dunning, and Roberto Tateo.
\newblock Supersymmetry and the spontaneous breakdown of {$\mathcal{PT}$}
  symmetry.
\newblock {\em J. Phys.}, A34:L391, 2001, [hep-th/0104119].

\bibitem{Bender:1992bk}
Carl~M. Bender and Alexander Turbiner.
\newblock Analytic continuation of eigenvalue problems.
\newblock {\em Phys. Lett.}, A173:442--446, 1993.

\bibitem{MR1627442}
Carl~M. Bender and Stefan Boettcher.
\newblock Real spectra in non-{H}ermitian {H}amiltonians having
  {$\mathcal{PT}$} symmetry.
\newblock {\em Phys. Rev. Lett.}, 80(24):5243--5246, 1998.

\bibitem{MR1686605}
Carl~M. Bender, Stefan Boettcher, and Peter~N. Meisinger.
\newblock {$\mathcal{P}\mathcal{T}$}-symmetric quantum mechanics.
\newblock {\em J. Math. Phys.}, 40(5):2201--2229, 1999.

\bibitem{MR1742959}
Eric Delabaere and Fr{\'e}d{\'e}ric Pham.
\newblock Eigenvalues of complex {H}amiltonians with {$\mathcal{P}\mathcal
  {T}$}-symmetry. {I}.
\newblock {\em Phys. Lett. A}, 250(1-3):25--28, 1998.

\bibitem{Dorey:1998pt}
Patrick Dorey and Roberto Tateo.
\newblock Anharmonic oscillators, the thermodynamic bethe ansatz, and nonlinear
  integral equations.
\newblock {\em J. Phys.}, A32:L419--L425, 1999, [hep-th/9812211].

\bibitem{MR1832065}
Vladimir~V. Bazhanov, Sergei~L. Lukyanov, and Alexander~B. Zamolodchikov.
\newblock Spectral determinants for {S}chr\"odinger equation and {${\bf
  Q}$}-operators of conformal field theory.
\newblock In {\em Proceedings of the Baxter Revolution in Mathematical Physics
  (Canberra, 2000)}, volume 102, pages 567--576, 2001.

\bibitem{Dorey:1999uk}
Patrick Dorey and Roberto Tateo.
\newblock On the relation between {S}tokes multipliers and the {$T$}-{$Q$}
  systems of conformal field theory.
\newblock {\em Nucl. Phys.}, B563:573--602, 1999, [hep-th/990621]9.

\bibitem{MR1832082}
J.~Suzuki.
\newblock Functional relations in {S}tokes multipliers---fun with {$x\sp
  6+\alpha x\sp 2$} potential.
\newblock In {\em Proceedings of the Baxter Revolution in Mathematical Physics
  (Canberra, 2000)}, volume 102, pages 1029--1047, 2001.

\bibitem{Dorey:2000kq}
Patrick~E. Dorey, Clare Dunning, and Roberto Tateo.
\newblock Ordinary differential equations and integrable quantum field
  theories.
\newblock 2000, [hep-th/0010148].

\bibitem{Dorey:2004ta}
Patrick Dorey, Clare Dunning, and Roberto Tateo.
\newblock Aspects of the {ODE / IM} correspondence.
\newblock 2004, [hep-th/0411069].

\bibitem{Griffiths}
David~J. Griffiths.
\newblock {\em Introduction to quantum mechanics}.
\newblock Prentice-Hall, Inc., 1994.

\bibitem{Heading:1962}
J~Heading.
\newblock {\em An Introduction to Phase-Integral Methods}.
\newblock Methuen and Co Ltd, London, 1962.

\bibitem{Berry:1972na}
Michael~V. Berry and K.~E. Mount.
\newblock Semiclassical approximations in wave mechanics.
\newblock {\em Rept. Prog. Phys.}, 35:315, 1972.

\bibitem{Voros83}
A.~Voros.
\newblock The return of the quartic oscillator. {T}he complex {WKB} method.
\newblock {\em Ann. Inst. Henri Poincar{$\acute{e}$}}, XXXIX(3):211--338, 1983.

\bibitem{BenderBerry:2001wk}
Carl~M Bender, Michael Berry, Peter~N Meisinger, Van~M Savage, and Mehmet
  Simsek.
\newblock Complex {WKB} analysis of energy-level degeneracies of
  non-{H}ermitian {H}amiltonians.
\newblock {\em Journal of Physics A: Mathematical and General}, 34(6):L31--L36,
  2001.

\bibitem{Dorey:unpub}
Patrick Dorey, Clare Dunning, and Roberto Tateo.
\newblock The {ODE/IM} correspondence.
\newblock 2007, [hep-th/0703066].

\bibitem{Dorey:unpub2}
Patrick Dorey, Clare Dunning, Anna Lishman, and Roberto Tateo.
\newblock to be published.

\end{thebibliography}

\end{document}